\documentclass[aps,prb,twocolumn,superscriptaddress,10pt]{revtex4-1}
\usepackage{amsmath,amssymb,amsfonts,upgreek}
\usepackage{color}
\usepackage{graphicx}
\usepackage{setspace}
\usepackage{bm}
\usepackage[export]{adjustbox}
\usepackage{comment}
\usepackage{hyperref}
\usepackage{dsfont}
\usepackage{braket}

\newcommand{\cfs}{crystal-field splitting}
\newcommand{\op}{orbital polarization}
\newcommand{\ttg}{\ensuremath{t_{2g}}}

\newcommand{\umit}{\ensuremath{U_{\textrm{\sc mit}}}}
\newcommand{\pd}{\ensuremath{p\textrm{-}d}}

\def\hn{\hat{n}}

\def\mp{m^\prime}
\def\hnmu{\hat{n}_{m\uparrow}}
\def\hnmd{\hat{n}_{m\downarrow}}

\def\hnpmd{\hat{n}_{m^\prime\downarrow}}
\def\spinup{\uparrow}
\def\spindown{\downarrow}

\begin{document}

\title{On the effect of charge self-consistency in DFT+DMFT calculations for complex transition metal oxides}

\author{Alexander Hampel}
\email{ahampel@flatironinstitute.org}
\affiliation{Materials Theory, ETH Z\"u{}rich, Wolfgang-Pauli-Strasse 27, 8093 Z\"u{}rich, Switzerland}
\affiliation{Center for Computational Quantum Physics, Flatiron Institute,
162 5th Avenue, New York, NY 10010, USA} 
\author{Sophie Beck}
\email{sophie.beck@mat.ethz.ch}
\affiliation{Materials Theory, ETH Z\"u{}rich, Wolfgang-Pauli-Strasse 27, 8093 Z\"u{}rich, Switzerland}
\author{Claude Ederer}
\email{claude.ederer@mat.ethz.ch}
\affiliation{Materials Theory, ETH Z\"u{}rich, Wolfgang-Pauli-Strasse 27, 8093 Z\"u{}rich, Switzerland}

\newcommand{\cvo}{CaVO$_3$}

\date{\today}

\begin{abstract}
We investigate the effect of charge self-consistency (CSC) in density functional theory plus dynamical mean-field theory (DFT+DMFT) calculations compared to simpler ``one-shot'' calculations for materials where interaction effects lead to a strong redistribution of electronic charges between different orbitals or between different sites.
We focus on two systems close to a metal-insulator transition,  for which the importance of CSC is currently not well understood.
Specifically, we analyze the strain-related orbital polarization in the correlated metal CaVO$_3$ and the spontaneous electronic charge disproportionation in the rare-earth nickelate LuNiO$_3$. In both cases, we find that the CSC treatment reduces the charge redistribution compared to cheaper one-shot calculations. 
However, while the MIT in CaVO$_3$ is only slightly shifted due to the reduced orbital polarization, the effect of the site polarization on the MIT in LuNiO$_3$ is more subtle.
Furthermore, we highlight the role of the double-counting correction in CSC calculations containing different inequivalent sites. 

\end{abstract}

\maketitle

\section{Introduction}

During recent years, the combination of density functional theory (DFT) and dynamical mean-field theory (DMFT) has become a widespread tool to calculate properties of so-called ``correlated materials'', i.e., materials where the strong Coulomb repulsion between electrons in partially filled $d$ or $f$ shells leads to effects that cannot easily be treated within effective non-interacting electron theories \cite{held2007}. 
The basic idea in combining DFT and DMFT is the assumption that for the relevant materials the electronic degrees of freedom can be separated into a ``weakly interacting'' part, for which a standard DFT treatment is adequate, and a ``correlated subspace'', which requires a more elaborate treatment of the electron-electron interaction. The latter leads, in general, to a redistribution of electrons within the correlated subspace compared to the DFT result. This change should then enter, in a self-consistent way, the effective potential felt by the weakly interacting electrons, which is achieved by iterating between DFT and DMFT steps. However, such a charge self-consistent (CSC) DFT+DMFT calculation leads to a higher computational cost compared to simpler ``one-shot'' (OS) calculations, where the charge rearrangement within the correlated subspace is neglected in the DFT calculation.

Thus, as DFT+DMFT develops further towards a standard {\it ab initio}-based computational method for materials science~\cite{Grieger_et_al:2012,Adler_et_al:2019}, it becomes essential to know in which cases it is possible to reduce the required computational effort by using more approximate variants of the method, e.g., by neglecting charge self-consistency.
While CSC DFT+DMFT calculations have become more common recently, the DFT+DMFT method has also been applied to larger and more complex systems, such as, e.g., oxide heterostructures,~\cite{Ishida:2009,Zhong:2015,Lechermann:2018,Beck:2019} defective systems,~\cite{Backes_et_al:2016,Sing:2017,Souto-Casares:2019} or large molecules~\cite{Jacob:2010,Turkowski:2012}.
Therefore, a detailed understanding of the effect of charge self-consistency is desirable in order to better assess in which cases a CSC calculation is crucial or, more importantly, in which circumstances a one-shot calculation is sufficient. 
Unfortunately, there are currently very few studies available that provide a systematic quantitative comparison between CSC and one-shot calculations.
It is the purpose of the present work to start filling this gap.

It can be assumed that charge self-consistency is particularly relevant for systems where correlation effects lead to a redistribution of electrons, e.g., for systems with charge-, and/or orbital-ordering. For example, existing studies of epitaxially strained SrVO$_3$ monolayers demonstrate a reduced orbital polarization in CSC calculations compared to OS~\cite{Bhandary:2016,Schuler_Aichhorn:2018}. 
Here, we therefore analyze the effect of charge self-consistency for two specific but representative cases. 
First, strained CaVO$_3$, where orbital polarization is considered relevant for a reported strain-induced metal-insulator transition~\cite{Gu_et_al:2013,Beck:2018}. And second, LuNiO$_3$, which is representative for a whole series of rare earth nickelates that exhibit a transition to a charge-ordered (or charge-disproportionated) insulating state, which is also strongly coupled to a structural distortion~\cite{catalano.review}.

Most previous work addressing the influence of charge self-consistency in DFT+DMFT calculations in transition metal (TM) oxides employed a so-called ``\pd{}''-model to define the correlated subspace,~\cite{Wang:2012,Dang:2014,Park:2014,Aichhorn:2009,Aichhorn:2011,Karolak:2010} i.e., using a basis of rather localized, atomic-like orbitals constructed from a broad energy window that includes the TM $d$ as well as all oxygen $p$ bands. 
This appears conceptually appealing, since a wider energy window corresponds to a larger, and thus more complete, basis set, and since the use of rather localized orbitals provides better justification for the DMFT assumption of a purely local self-energy and Coulomb interaction \cite{Aichhorn:2011}. 
On the other hand, this also increases the computational load compared to using a minimal correlated subspace of ``frontier'' orbitals, corresponding to only a narrow energy window around the Fermi level.
In TM oxides, the latter typically includes either $t_{2g}$ or $e_g$ bands.

In the present work, we focus on DFT+DMFT calculations that employ such a minimal correlated subspace, corresponding to only a small number of near-Fermi-surface bands. These are expressed in a localized basis through a suitable transformation in terms of Wannier functions~\cite{PhysRevB.74.125120}.
By including only the minimal number of orbitals needed to describe the dominant low-energy physics within DMFT, this scheme requires a comparatively small computational cost. Furthermore, it often allows for an intuitive interpretation of Wannier occupations in terms of formal charge states, since the corresponding Wannier functions include the hybridization with the oxygen $p$ states as ``tails'' located on the oxygen sites.

Another critical point arising in DFT+DMFT calculations using a \pd{}-type orbital subspace, is that the physically very important charge transfer energy, $\Delta_{\pd{}}$, which describes the energy difference between oxygen $p$ and transition metal $d$ states, effectively becomes controlled by the so-called double-counting correction \cite{Wang:2012,Karolak:2010,Solovyev/Terakura:1998}. The latter is required to account for the electron-electron interaction within the correlated subspace that is already included on the DFT level, and is notoriously ill-defined \cite{Kotliar:2006}.
Different expressions to account for the double counting (DC) have been suggested \cite{Karolak:2010,Haule:2015_exactDC}, but in some cases the double-counting needs to be adjusted manually, in order to obtain satisfactory results~\cite{Dang:2014,Wang:2012}.
It was shown that CSC calculations for such \pd{}-type calculations produce essentially the same spectral properties as OS calculations, if one tunes the DC correction to yield the same $d$-state occupancy~\cite{Wang:2012}. 
It is, however, not clear a priori, that more complex observables, e.g., the total energy, need to agree within both approaches.
We note that the use of a minimal correlated subspace avoids the problem that the DC correction critically affects the important charge transfer energy, because charge-neutrality between DFT and DMFT is ensured, and thus the DC potential shift can be absorbed in the chemical potential in DMFT~\cite{Schuler_Aichhorn:2018}.
However, as we show in the following, the DC correction can still have a strong effect for systems with multiple inequivalent correlated sites.

In the next section (Sec.~\ref{sec:theory}), we provide a detailed description of the DFT+DMFT method as used in this work, specifying also all important computational parameters. We then discuss the two specific cases of CaVO$_3$ and LuNiO$_3$ in Sec.~\ref{sec:results}, where we also provide a brief introduction in the relevant physical background for each of these materials. Finally, our main conclusions are summarized in Sec.~\ref{sec:summary}.

\section{Theoretical framework}
\label{sec:theory}

\subsection{DFT calculations}

Structural relaxations for CaVO$_3$ within the 20 atom unit cell in $Pbnm$ space group symmetry are performed using the \textsc{Quantum~ESPRESSO} package~\cite{Giannozzi_et_al:2009}.
We employ scalar-relativistic ultrasoft pseudopotentials, with the $3s$ and $3p$ semicore states included in the valence for both V and Ca,
together with the exchange-correlation functional according to Perdew, Burke, and Ernzerhof~\cite{Perdew:1996iq}.
Cell parameters and internal coordinates are relaxed until all force components are smaller than 0.1 mRy/$a_0$ ($a_0$: Bohr radius) and all components of the stress tensor are smaller than 0.1 kbar. The plane-wave energy cutoff is set to 70~Ry for the wavefunctions and 840~Ry for the charge density. A $6 \times 6 \times 4$ Monkhorst-Pack $k$-point grid is used to sample the Brillouin zone, and the Methfessel-Paxton scheme with a smearing parameter of 0.02 Ry is used to broaden electron occupations.
For the calculation of epitaxially strained CaVO$_3$, the in-plane lattice parameters are increased by 4\% and kept fixed, while the $c$-component of the cell and all atomic positions are relaxed.

All DFT calculations for LuNiO$_3$ as well as the DFT parts of all our CSC DFT+DMFT calculations are performed using the projector augmented wave (PAW) method~\cite{Blochl:1994dx}, implemented in the ``Vienna Ab initio Simulation Package''(VASP)~\cite{Kresse:1993bz,Kresse:1996kl,Kresse:1999dk}, and also using the exchange correlation functional according to Perdew, Burke, and Ernzerhof~\cite{Perdew:1996iq}.
For Ni, we use the PAW potential where the 3$p$ semi-core states are included as valence electrons, while for Lu, we use the PAW potential corresponding to a $3+$ valence state with $f$-electrons frozen into the core. 
For the CaVO$_3$ calculations with VASP, we use the PAW potentials where the $s$ and $p$ semi-core states are included as valence electrons for both Ca and V. 
Furthermore, a $k$-point mesh with $9 \times 9 \times 7$ grid points along the three reciprocal lattice directions is used and a plane wave energy cut-off of 550~eV is chosen for LuNiO$_3$ and 600~eV for CaVO$_3$.
The structure of LuNiO$_3$ is fully relaxed within $Pbnm$ symmetry, both internal and lattice parameters, until the forces acting on all atoms are smaller than $10^{-4}$ eV/\r{A}. 

\subsection{DFT+DMFT calculations}

\subsubsection{Construction of the correlated subspace}
In the DFT+DMFT method, the Kohn-Sham (KS) Hamiltonian within the chosen energy window is mapped onto a basis of localized states, spanning the correlated subspace $\mathcal{C}$, then a local Coulomb interaction is added, and the resulting Hubbard-like lattice Hamiltonian is solved via the DMFT approximation~\cite{Georges:1996,held2007}.
Without feedback to the DFT part, this corresponds to a OS calculation. To perform CSC calculations, one computes a correction to the charge density, $\Delta \rho = \rho^\text{DMFT} - \rho^\text{DFT}$, which is then passed back to the DFT code (here VASP) to calculate new KS wave-functions and hence, update the correlated subspace~\cite{Amadon:2008,Lechermann:2018}.
In a fully CSC calculation, this is repeated until $\Delta \rho$ does not change compared to the previous iteration.

For the DMFT calculation, the electronic degrees of freedom within the chosen energy window are described via the interacting lattice Green's function:
\begin{align}
    \hat{G}(\mathbf{k}, i \omega_n) = \left[ (i \omega_n + \mu) \mathds{1} - \hat{H}_{\text{KS}}(\mathbf{k}) - \hat{\Sigma}(\mathbf{k}, i \omega_n) \right]^{-1}
\label{eq:Glat}
\end{align}
where $\mu$ is the chemical potential and $\hat{H}_{\text{KS}}(\mathbf{k})$ is the Kohn-Sham (KS) Hamiltonian.
The lattice self-energy $\hat{\Sigma}(\mathbf{k}, i \omega_n)$ is obtained by solving the effective DMFT impurity problem (see next sub-section).

The lattice Green's function in Eq.~\eqref{eq:Glat} is expressed in the KS (Bloch) basis.
To achieve the up/down-folding between the quantities defined within the correlated subspace and the Green's function in the KS basis,
\begin{align}
    {G}^{\mathcal{C}}_{L L'} (i \omega_n) = \sum_{k, \nu
    \nu'} {P}_{L \nu}(\mathbf{k}) \ {G}_{\nu \nu'}(\mathbf{k}, i \omega_n) \  {P}^{*}_{\nu' L'}(\mathbf{k}) \ ,
\end{align}
projector functions ${P}_{L \nu}(\mathbf{k})$ are introduced.
The projector functions are defined as projections of the KS eigenstates $\ket{\Psi_{\nu \mathbf{k}}}$ onto localized orbitals 

$\ket{\chi_L}$, ${P}_{L \nu}(\mathbf{k}) \equiv \braket{\chi_L | \Psi_{\nu \mathbf{k}}}$.
Here, $L$ serves as compound index for all local quantum numbers (site, orbital, and spin-character).

In our VASP-based OS and CSC calculations, the local basis functions $\ket{\chi_L}$ are constructed from projection to localized orbitals (PLO)~\cite{Amadon:2008,Schuler_Aichhorn:2018}.
To construct optimal projector functions, we apply the scheme introduced in Ref.~\onlinecite{Schuler_Aichhorn:2018}, choosing a linear combination of the PAW partial wave augmentation channels that maximizes the overlap with the KS states inside a chosen energy window, which matches that of the correlated subspace $\mathcal{C}$.
We use initial projections on \ttg{}- or $e_g$-like orbitals within the energy window of the correlated subspace $\mathcal{C}$. The resulting projectors $\tilde P_{L, \nu} (\mathbf{k})$ are in general not orthogonal to each other, and need to be orthonormalized:
    \begin{equation}
    O_{L L'}(\mathbf{k}) = \sum_{\nu} {\tilde P}_{L \nu}(\mathbf{k}) {\tilde P}^*_{\nu L'}(\mathbf{k}) \quad ,
\end{equation}
\begin{equation}
    P_{L \nu}(\mathbf{k}) = \sum_{L'} [O^{-1/2}(\mathbf{k})]_{L L'} \  {\tilde P}_{L' \nu}(\mathbf{k}) \quad ,
\end{equation}
to define an orthonormal set of PLO-based Wannier functions describing the correlated subspace $\mathcal{C}$. The orthonormalization of these PLO-based Wannier functions, as well as the whole DFT+DMFT self-consistency cycle has been implemented using the \textsc{TRIQS/DFTTools} software package~\cite{aichhorn_dfttools_2016,parcollet_triqs_2015}. 

The projectors ${P}_{L \nu}(\mathbf{k})$ are updated after each DMFT cycle from the new KS states. Thereby, the energy window defining the correlated subspace is kept fixed, while monitoring that the change in the KS energies due to the charge density correction does not move the relevant bands outside of this energy window. The observed change of the KS eigenvalues is relatively small for all cases considered in this work, e.g., the maximum bandwidth reduction in LuNiO$_3$ is smaller than $\sim5$\%. 

We note that the strong octahedral rotations present within $Pbnm$ symmetry lead to large off-diagonal crystal-field terms in the KS Hamiltonian, and the non-interacting Green's function for the effective impurity problem is no longer diagonal.
Since this can induce a severe sign problem in the quantum Monte Carlo (QMC) method~\cite{Gull:2011} used to solve the effective impurity problem (see next sub-section), we perform a local unitary transformation of each impurity Green's function after the down- respectively before the up-folding, which diagonalizes the initial non-interacting local Hamiltonian on each site transforming the system into the crystal field basis.
We note that the transformation is optimized in the first CSC cycle, and is kept fixed in consecutive iterations to maintain a consistent orbital basis.

For \cvo{} we also perform OS DFT+DMFT calculations based on the electronic structure obtained with \textsc{Quantum~ESPRESSO}.
In this case, the low-energy tight-binding Hamiltonian, used as input for the OS DMFT calculation, is formulated in the basis of maximally localized Wannier functions (MLWFs)~\cite{Marzari_et_al:2012} using the \textsc{Wannier90} code~\cite{Mostofi_et_al:2014}.
Note that the PLO basis functions used in our VASP-based DFT+DMFT calculations essentially correspond to the initial Wannier functions constructed by \textsc{Wannier90} before the spread minimization, which are based on orthogonalized projections of (pseudo-) atomic orbitals on the Bloch bands~\cite{Mostofi_et_al:2014}.

The code used for all DFT+DMFT calculations in this paper is publicly available on github~\cite{soliDMFT}.

\subsubsection{Solving the impurity problem}

For both CaVO$_3$ (\ttg{} subspace) and LuNiO$_3$ ($e_g$ subspace) the effective impurity problem within the DMFT cycle is solved with a continuous-time QMC hybridization-expansion solver~\cite{Gull:2011} implemented in \textsc{TRIQS/cthyb}~\cite{Seth2016274}, taking into account all off-diagonal elements of the local Green's function in the crystal-field basis.
For each impurity we add a local Coulomb interaction in the form of the Hubbard-Kanamori Hamiltonian~\cite{vaugier2012},
\begin{align}
\begin{split}
\hat{H}_{\mathrm{int}}\,&=\,U\sum_m \hnmu\hnmd\,+\,(U-2J)\sum_{m\neq\mp} \hnmu\hnpmd \\
&+(U-3J) \sum_{m<\mp,\sigma} \hn_{m\sigma}\hn_{\mp\sigma} \\ &+ J\, \sum_{m\neq\mp} \hat{c}^{\dagger}_{m\spinup}\hat{c}^{\dagger}_{m\spindown}\,\hat{c}_{\mp\spindown}\hat{c}_{\mp\spinup}
-J\,\sum_{m\neq\mp} \hat{c}^{\dagger}_{m\spinup}\hat{c}_{m\spindown}\,\hat{c}^{\dagger}_{\mp\spindown}\hat{c}_{\mp\spinup} \ ,
\end{split}
\label{eq:ham_kanamori}
\end{align}
including all spin-flip and pair-hopping terms.
Here, the operator $\hat{c}^{\dagger}_{m\sigma}$ creates an electron in the atom-centered Wannier orbitals of type $m$ and spin $\sigma$.
The interaction parameters are given by the local intra-orbital Coulomb repulsion $U$, and the Hund's coupling $J$.
To reduce the QMC noise in the high-frequency regime of the impurity self-energy $\Sigma_\text{imp}$ and $G_\text{imp}$, we represent both quantities in the Legendre basis~\cite{boehnke:2011} and sample the Legendre coefficients $G_l$ directly within the \textsc{TRIQS/cthyb} solver.

\subsubsection{Double counting correction}\label{par:DC}

To correct the electron-electron interaction within the correlated subspace already accounted for within VASP, we use the fully-localized limit DC correction scheme~\cite{Solovyev:1994,anisimov1997}. Specifically, we use the parameterization given in Ref.~\onlinecite{held2007} for the DC potential,
\begin{align}
\label{eq:dcimp}
\Sigma_{dc,\alpha}^\text{imp} = \bar{U}   (n_{\alpha}-\frac{1}{2}) \quad ,
\end{align}
where $n_{\alpha}$ is the occupation of impurity site $\alpha$, and the average Coulomb interaction between $M$ orbitals, $\bar{U}$, is defined as~\cite{held2007}
\begin{align}
\label{eq:barU}
    \bar{U} = \frac{U+(M-1)(U-2J)+(M-1)(U-3J)}{2M-1} \ .
\end{align}
The potential shift of Eq.~\eqref{eq:dcimp} is added to the impurity self-energy; its form is directly tailored to the Hubbard-Kanamori interaction Hamiltonian in Eq.~\eqref{eq:ham_kanamori} for a \ttg{}- or $e_g$-model resulting from an octahedral crystal-field environment of $M$ interacting orbitals ($M=3$ for \cvo{} and $M=2$ for LuNiO$_3$).

In this work, we draw particular attention on how the occupations $n_\alpha$ used for the DC correction are evaluated, i.e., whether they correspond to:
a) the occupations of the Wannier functions as obtained from DFT, or b) the occupations corresponding to the impurity Green's function $G_\text{imp}$ calculated by the QMC solver within the DMFT step.
It can be misleading to assume that these quantities are the same, even within a CSC calculation.
Indeed, when the system is in a charge-ordered phase, such as, e.g., in heterostructures or nickelates, or in any other case with several inequivalent impurity problems, different impurities can exchange charge within the DMFT loop, potentially leading to drastic changes of the local occupations compared to the ones calculated within the DFT step.
In principle, only the occupations evaluated for the impurity problem within DMFT that are used to define the charge density correction, have physical meaning within a CSC DFT+DMFT calculation.
By contrast, the occupations obtained in the DFT part, by filling up the lowest energy KS states, do not correspond to the charge density that is used to evaluate the Kohn-Sham potential in a CSC calculation. 
However, in the case of a OS DFT+DMFT calculation, the question of whether to use DFT or DMFT occupations for the DC correction is ambiguous.
An informal (and perhaps unrepresentative) community survey conducted by us, has shown that both variants are currently used in different studies.
Here, we show that in certain systems the question of how to extract $n_{\alpha}$ can have a strong influence on the results, and that one should be aware of this issue when evaluating the DC correction.

\subsubsection{Calculation of observables}

From the imaginary-time Green's function, we calculate the spectral weight around the Fermi level, \mbox{$\bar{A}(\omega=0) = - \frac{\beta}{\pi} {\rm Tr} \ G^{\mathcal{C}}_{L L'} \left( \beta/2 \right)$}, which indicates whether the system is metallic \mbox{($\bar{A}(0) > 0$)} or insulating \mbox{($\bar{A}(0) \approx 0$)}~\cite{Fuchs:2011}.
For $T=0$ ($\beta \rightarrow \infty$), $\bar{A}$ is identical to the spectral function at $\omega=0$. For finite temperatures, it represents a weighted average around $\omega=0$ with a width of $\sim k_\text{B}T$~\cite{Fuchs:2011}.The full real-frequency spectral function $A(\omega)$ is obtained via analytic continuation using the maximum entropy method~\cite{Jarrel:2010}.
The on-site density matrix can be obtained directly from the local Matsubara Green's function as \mbox{$n_{L L'} = \frac{1}{\beta} \sum_{\omega_n}  G^{\mathcal{C}}_{L L'} (i \omega_n)$}.

To extract the total energy of the system we use the following formula~\cite{PhysRevB.74.125120}:
\begin{align}
\begin{split}
E_{\text{DFT+DMFT}} &= E_{\text{DFT}}[\rho] \\     &- \frac{1}{N_k} \sum_{\nu \in \mathcal{C} ,\mathbf{k}} \epsilon_{\nu,\mathbf{k}}^{\text{KS}} \ f_{\nu \mathbf{k}} + \langle \hat{H}_{\text{KS}} \rangle_{\text{DMFT}} \\     & + \langle \hat{H}_{\text{int}} \rangle_{\text{DMFT}} - E_{\text{DC}}^\text{imp} \quad ,
\end{split}
\label{eq:dmft-dft-tot-en-2}
\end{align}
where $\epsilon_{\nu,\mathbf{k}}^{\text{KS}}$ are the KS eigenvalues with corresponding weights $f_{\nu \mathbf{k}}$ within the correlated subspace $\mathcal{C}$, and $\langle \cdot \rangle_{\text{DMFT}}$ denotes quantities evaluated from the DMFT solution. The interaction energy $\langle \hat{H}_{\text{int}} \rangle_{\text{DMFT}}$ is calculated using the Galitskii-Migdal formula~\cite{abrikosov2012methods,galitskii1958}, and the last term in Eq.~\eqref{eq:dmft-dft-tot-en-2} subtracts the DC energy. 
To ensure high accuracy, we sample the total energy over a minimum of additional 60 converged DMFT iterations after the CSC DFT+DMFT loop is already converged.
Convergence is reached when the standard error of the site occupation during the last 10 DFT+DMFT loops is smaller than $1.5 \times 10^{-3}$. This way, we achieve an accuracy in the total energy of $<5$\,meV. All DMFT calculations are performed for $\beta=40$ eV$^{-1}$, which corresponds to a temperature of 290~K.

\section{Materials \& Results}
\label{sec:results}

\begin{figure*}
    \centering
    \includegraphics[width=0.9\linewidth]{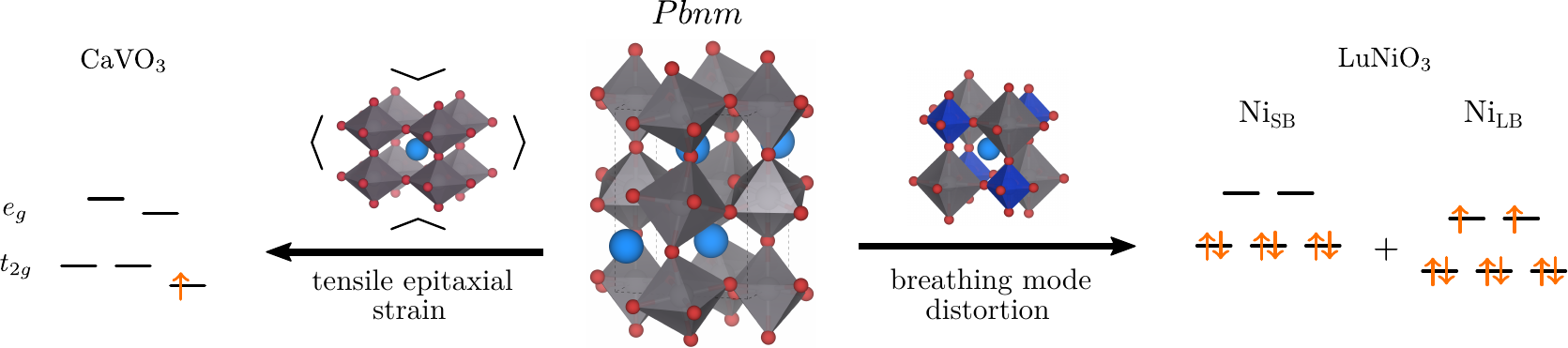}
    \caption{Exemplary $Pbnm$ crystal structure, as well as the main distortion modes relevant for the discussion of \cvo{} and LuNiO$_3$, respectively, i.e., tetragonal strain (left) and octahedral breathing mode distortion (right).
    The $d$-orbital energy levels that result from these distortions and the nominal occupations for each compound are also depicted.
    Note that for simplicity we omitted octahedral rotations in the simplified lattice structures and the corresponding $d$-level crystal-field splittings, and that the local spin direction fluctuates in the paramagnetic case.
    }
    \label{fig:structure}
\end{figure*}

To analyze the effect of CSC within DFT+DMFT, we study two representative examples of TM oxides with different levels of complexity. First, we consider the case of unstrained and strained \cvo{}.
While in the former case this material is a correlated metal~\cite{Nekrasov:2005,Pavarini_et_al:2004}, it has recently been demonstrated that tensile epitaxial strain leads to a transition towards the Mott insulating state within OS DFT+DMFT calculations~\cite{Beck:2018}.
An important aspect in this transition is the strain-induced crystal-field splitting between the partially filled $t_{2g}$ orbitals, leading to a strong orbital polarization, and thus a local charge redistribution, which can potentially affect the result of a CSC compared to a OS DFT+DMFT calculation. 
However, in \cvo{}, all correlated sites are symmetry-equivalent and thus the DC correction is irrelevant when using a minimal ``$t_{2g}$-only'' correlated subspace. 

Second, we consider the rare earth nickelate LuNiO$_3$, which exhibits a complex interplay between a specific structural distortion and an associated charge ordering, resulting in a transition from a paramagnetic metallic towards an also paramagnetic but insulating phase~\cite{catalano.review}. In previous works, it was shown that DFT+DMFT is well suited to describe this correlated insulating state~\cite{Park:2012hg,Park:2014,Subedi:2015en}. Since two symmetry-inequivalent types of Ni sites appear in the insulating state, this case allows to analyze the effect of a site-dependent DC correction within CSC DFT+DMFT calculations.

Both materials, \cvo{} and LuNiO$_3$, exhibit a distorted perovskite structure with $Pbnm$ space group (in the case of LuNiO$_3$ this corresponds to the high symmetry metallic phase). The corresponding unit cell contains four TM atoms surrounded by edge-connected oxygen octahedra, that are tilted and rotated around the Cartesian axes, corresponding to the so-called GdFeO$_3$-type distortion ($a^-a^-c^+$ tilt system in Glazer notation), as depicted in Fig.~\ref{fig:structure}.
The $d$-levels of the TM ions are split into $e_g$ and $t_{2g}$ manifolds by the octahedral crystal field, and the remaining degeneracies can be further lifted by additional distortions of the oxygen octahedra (also shown schematically in Fig.~\ref{fig:structure}).

\subsection{\cvo{} - orbital polarization}
\label{sec:cvo}

As stated above, bulk \cvo{} is a moderately correlated metal with weak \op{} that can undergo a transition to the Mott-insulating state under tensile epitaxial strain or in ultra-thin films~\cite{Beck:2018,Gu_et_al:2013,Mcnally:2019}.
As has been pointed out in Ref.~\onlinecite{Pavarini:2005}, the \op{} resulting from the orthorhombic distortion of the perovskite structure (related to the tilts and rotations of the octahedral network) is an important factor in the metal-insulator transition (MIT).
Several examples suggest that by an appropriate tuning of the bandwidth and the \cfs{} via, for example, strain or dimensional confinement, the resulting charge redistribution enhances the \op{}, eventually leading to a MIT~\cite{Gu_et_al:2013,Sclauzero/Dymkowski/Ederer:2016,Beck:2018}.
For example, as depicted in Fig.~\ref{fig:structure}, tensile epitaxial strain will lift the degeneracy of the $t_{2g}$-states, lowering the energy of one orbital compared to the other two.
Since the \op{} in \cvo{} can be seen as a measure for the likelihood of the Mott-insulating state, it is clear that describing this quantity accurately is essential for the success of the chosen method.

\begin{figure}
    \centering
    \includegraphics[width=\linewidth]{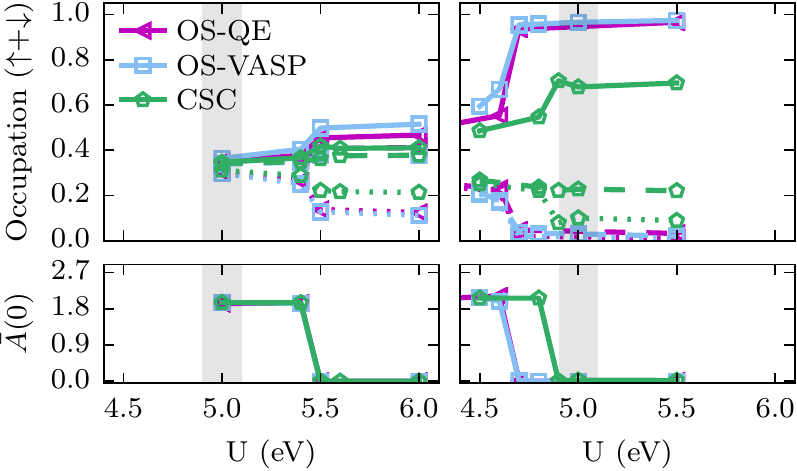}
    \caption{DFT+DMFT results obtained from OS and CSC calculations with VASP (PLO basis), compared to OS calculations using \textsc{Quantum~ESPRESSO} (QE, MLWF basis), for bulk (left) and strained (right) CaVO$_3$.
    Top panels: Orbitally-resolved occupations as a function of the interaction parameter $U$. Bottom panels: averaged spectral weight at the Fermi level, $\bar A(0)$.
    }
    \label{fig:dmft_cvo}
\end{figure}

As described in Sec.~\ref{sec:theory}, we perform DFT+DMFT calculations for the bulk structure of \cvo{} using three different schemes, i.e., OS calculations using either MLWFs (magenta line in Fig.~\ref{fig:dmft_cvo}) or PLOs (blue lines in Fig.~\ref{fig:dmft_cvo}) to represent the correlated subspace, as well as CSC calculations using PLOs (green lines in Fig.~\ref{fig:dmft_cvo}).
From this we obtain the orbital occupations and spectral weight at the Fermi level, shown in Fig.~\ref{fig:dmft_cvo}, as a function of the Coulomb interaction parameter $U$.
In all cases, the spectral weight is finite for small values of $U$, where the system is metallic, and then becomes zero in the insulating phase for large $U$, with a rather sharp transition at $U_\text{MIT}$.
For the unstrained bulk system, all three approaches give identical results for the spectral weight as function of $U$, with a critical value of $U_\text{MIT}$=5.5\,eV. 
Thus, at $U \approx 5$ eV, which is typically considered as realistic value for $3d^1$ transition metal oxides~\cite{Pavarini_et_al:2004}, we find a finite weight corresponding to metallic behaviour, in agreement with experimental observations. 
This shows that the obtained results do not depend on details of the implementation, such as small differences in the basis used to represent the correlated subspace.

From the occupations shown in Fig.~\ref{fig:dmft_cvo} (top left), it can be seen that the orbital polarization is weak in the metallic regime, but is significantly enhanced above \umit{}, where the occcupation of one orbital is decreased compared to the other two orbitals. This is in line with the crystal-field splitting of the bulk on-site Wannier energies, where one orbital is energetically higher than the other two, with only a small difference between the latter~\cite{Beck:2018}. 
Here, the two different OS calculations agree extremely well, while the orbital polarization is slightly reduced in the CSC calculation, however with no apparent effect on the predicted \umit{}.

Under 4\% tensile strain (right panels in Fig.~\ref{fig:dmft_cvo}), the MIT is shifted to lower $U$ values, below the realistic value of $U \approx 5$\,eV.
Here, both the MLWF- and PLO-type OS calculations agree within the accuracy of the method, and give exactly the same value for the critical interaction parameter of $U_\text{MIT}=4.7$\,eV.
The CSC calculation, however, places the MIT at a slightly higher value of $U_\text{MIT}=4.9$ eV.

An even stronger difference between OS and CSC calculations can be seen in the orbital polarization, which is generally strongly enhanced compared to the unstrained case, due to a large strain-induced crystal-field splitting~\cite{Sclauzero/Dymkowski/Ederer:2016,Beck:2018} (see Fig.~\ref{fig:structure}).

Within the OS calculations, both PLO- and MLWF-based, we find that in the insulating regime two orbitals become completely empty, while the third one is essentially fully occupied by a single electron, i.e., the system exhibits full orbital polarization.
In the CSC calculation this orbital polarization is significantly reduced, with a maximal occupation of $\sim0.7$ in the preferential orbital.
The crystal-field-induced orbital polarization, enhanced by electronic interaction effects, has previously been suggested to be an important factor supporting the insulating phase~\cite{Pavarini_et_al:2004}, since the resulting effective half-filling of only one orbital promotes the MIT as opposed to fractional occupation of three degenerate levels.
This is consistent with our results, since the lower orbital polarization in the CSC calculation correlates with a higher \umit{} compared to the OS case.
\begin{figure}[t]
    \centering
	\includegraphics[width=\linewidth,]{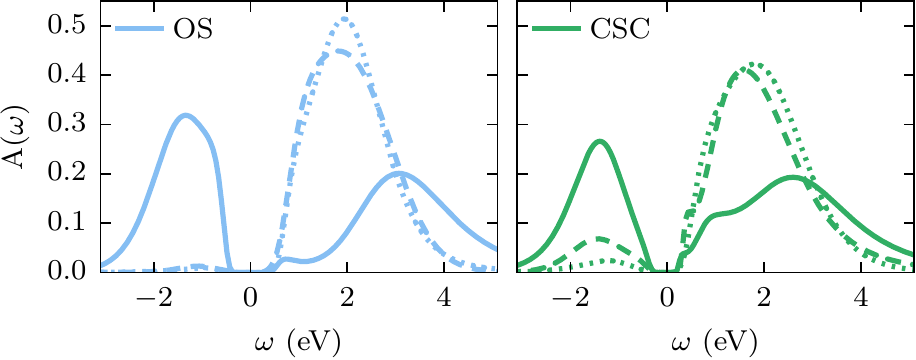}
	\caption{Orbitally-resolved spectral functions for CaVO$_3$ under 4\,\% tensile epitaxial strain, obtained from (PLO-based) OS (left) and CSC (right) DFT+DMFT calculations.}
	\label{fig:spec_cvo}
\end{figure}
To illustrate the difference between OS and CSC calculations in the strained case, we plot the spectral function $A(\omega)$ at $U=5.0$\,eV for both cases in Fig.~\ref{fig:spec_cvo}.
Here, the three different line-styles correspond to the three different \ttg{}-like orbitals.
As discussed previously, in the OS calculation one of the orbitals is essentially completely occupied, while the remaining two are empty.
In contrast to this, the CSC calculation shows a correlation-induced charge redistribution from the occupied orbital to the previously empty orbitals.
Furthermore, comparing the gap sizes of both cases, it is clearly visible that in the CSC case the gap is reduced compared to OS, similar to what has been reported in earlier studies on SrVO$_3$~\cite{Bhandary:2016}.

Overall, we conclude that charge self-consistency plays only a minor role for systems with weak or vanishing orbital polarization, where the corresponding charge redistribution obtained within DMFT compared to the DFT calculation is small. In contrast, for systems with strong differences in orbital occupations, the OS calculation can lead to an overestimation of the orbital polarization, which in turn can affect the tendency of the system to undergo a MIT.
While the effect on \umit{} is not too strong in the present case, the corresponding differences in spectral properties can be more pronounced.
Nevertheless, it appears that for the present case, OS calculations can at least give reliable qualitative information about the overall system behavior, such as, e.g., the effect of tensile epitaxial strain on \umit{}, favoring the insulating state.

Furthermore, we note that in our calculations using frontier orbitals, we find very good agreement between the PLO- and MLWF-based method, both in the spectral properties and for the orbital occupations.
This is in contrast to previous studies, reporting that projector-based methods require a larger U in \pd{} models due to larger hybridization effects~\cite{Dang:2014}. 

\subsection{LuNiO$_3$ --- charge-ordering and structural energetics}

The second case that we analyze is LuNiO$_3$.
This material belongs to the family of rare-earth nickelates, $R$NiO$_3$, where $R$ can be any rare-earth ion ranging from Lu to Pr, including Y.
All members of the series exhibit a MIT, which is accompanied by a structural transition, lowering the space group symmetry from $Pbnm$ to $P2_1/n$.
The corresponding structural distortion results in a three dimensional checkerboard-like arrangement of long bond (LB) and short bond (SB) NiO$_6$ octahedra, referred to as breathing mode distortion~\cite{Medarde2008}, and schematically shown on the right side of Fig.~\ref{fig:structure}.
Recent theoretical work indicates that this transition is related to an electronic instability towards spontaneous charge disproportionation on the Ni sites, which couples to the breathing mode, leading to a first-order coupled structural-electronic transition into a paramagnetic charge-disproportionated insulator (CDI)~\cite{peil:2019,hampel:2019}.
Furthermore, the choice of the $R$ site cation determines the degree of octahedral rotations in the corresponding high symmetry $Pbnm$ structure, and thus the bandwidth.
The latter then controls how close the system is to the electronic instability, driving trends across the series~\cite{peil:2019,hampel:2019,Mercy2017,Varignon:2017is,hampel2017}.

Here, we use the case of LuNiO$_3$ to analyze if, and how, the charge disproportionation, as a specific example for charge-ordering phenomena in general, is affected by the inclusion of charge self-consistency in DFT+DMFT. 
Earlier studies by \citet{Park2014short} also investigated the effect of CSC and DC for LuNiO$_3$ using a \pd{}-type subspace.
They found only a small effect due to CSC on total energy calculations, but had to adjust the DC correction to obtain a stable finite equilibrium breathing mode distortion.
Here, we use only the two $e_g$-like frontier orbitals per Ni site for our DFT+DMFT calculations.
As shown in Ref.~\onlinecite{Subedi:2015en}, the electronic instability towards charge disproportionation and the resulting \emph{site-selective Mott transition}~\cite{Park:2012hg} occurring in the paramagnetic state is well described within DFT+DMFT using such a minimal subspace. 

\citet{Subedi:2015en} found that the CDI state emerges in the frontier $e_g$ model for nickelates for rather large values of the Hund's coupling $J$, and is very sensitive to its variation. The fact that the Hund's coupling $J$ is the critical ingredient in the occurence of the CDI state was first proposed by \citet{Mazin:2007}. They showed in an atomic picture that when $U-3J$ becomes small and is overcome by the potential energy difference between the Ni sites, $\Delta_s$, which results from the breathing mode distortion and the charge disproportionation, the CDI state is favored. 
This regime is accessible in systems with small or negative charge-transfer gap, which results in a strong screening of the Coulomb interaction in the effective $d$ bands, whereas the Hund's coupling is less sensitive to screening~\cite{Mazin:2007}. 
A strong screening of $U$ in nickelates has been confirmed by recent studies using the constrained random phase approximation (cRPA)~\cite{Seth:2017,hampel:2019}. Moreover, in Ref.~\onlinecite{Isidori:2019} it is shown, that such a CDI regime for small or negative $U-3J$ is also accessible in a general three orbital Hubbard model, and is thus not limited to nickelate systems.

To isolate the effect of the structural breathing mode distortion on the electronic charge disproportionation and the total energy of the system, we distinguish the various structural distortions present in LuNiO$_3$ by employing a symmetry-based mode decomposition~\cite{PerezMato:2010ix}, as outlined in Refs.~\onlinecite{Balachandran:2013cg,hampel2017,hampel:2019}. 
This allows to add the breathing mode distortion, with symmetry label $R_1^+$, on top of the relaxed $Pbnm$ structure, and systematically vary its amplitude without changing any other parameter of the unit cell.
We use the software ISODISTORT \cite{Campbell:2006} to perform the mode decomposition. 

\subsubsection{Results for fixed structure}

\begin{figure}[t]
    \centering
    \includegraphics[width=0.8\linewidth]{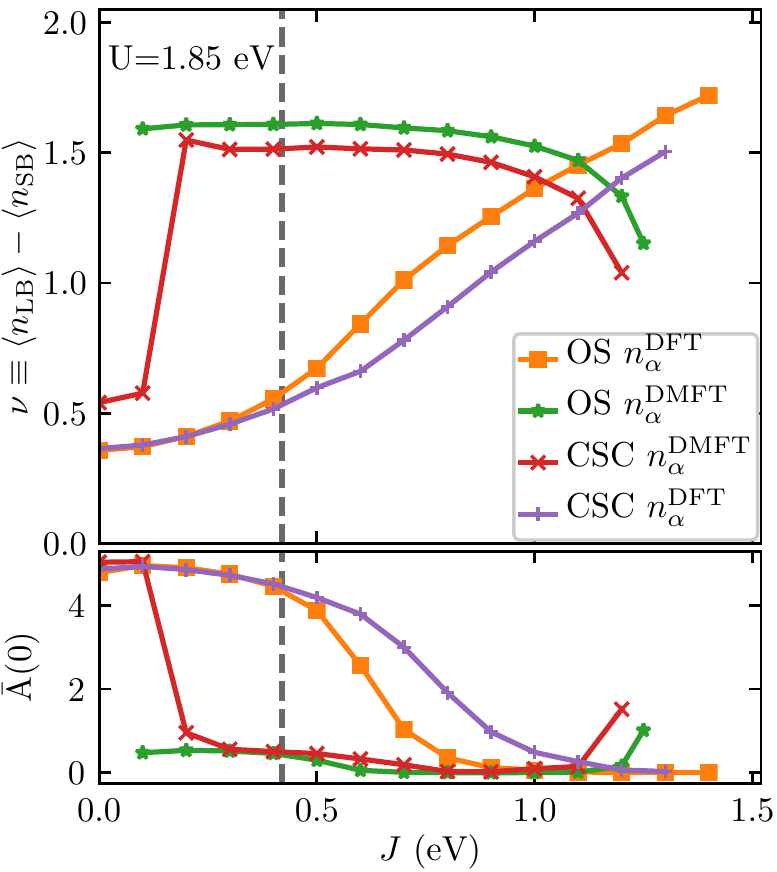}
    \caption{Results of different DFT+DMFT calculations for LuNiO$_3$ using the experimental $R_1^+$ amplitude, $U=1.85$\,eV, and varying $J$. CSC and OS calculations are labeled accordingly. For calculations labeled $n_\alpha^\text{DFT}$ ($n_{\alpha}^\text{DMFT}$) the DFT (DMFT) occupations have been used to evaluate the DC corrcetion. The dashed vertical line marks the cRPA value of $J$~\cite{hampel:2019}. Top: charge disproportionation $\nu$; bottom: corresponding spectral weight at the Fermi level.}
    \label{fig:lno_dc_comparison}
\end{figure}

First, we calculate the properties of LuNiO$_3$ for a fixed structure, using the experimentally observed breathing mode amplitude, $R_1^+=0.075$~\r{A}~\cite{Alonso:2001bs}, thereby varying the strength of the Hund's coupling $J$. As discussed above and shown in Ref.~\onlinecite{Subedi:2015en}, the charge disproportionation and the resulting MIT depend sensitively on $J$, which thus allows us to critically examine the influence of CSC on the most crucial system properties. 
We use a fixed $U$ value of 1.85~eV, which corresponds to the value calculated for LuNiO$_3$ using the cRPA~\cite{hampel:2019,Seth_Georges:2017}.
The results are depicted in Fig.~\ref{fig:lno_dc_comparison}, where in the top panel the charge disproportionation, $\nu \equiv \langle n_{\text{LB}} \rangle - \langle n_{\text{SB}} \rangle$, i.e. the difference of the $e_g$ occupation between the LB and SB Ni sites, is shown as function of $J$. The bottom panel shows the corresponding value for $\bar{A}(0)$, indicating whether the system is metallic or insulating. The dashed vertical line corresponds to the $J$ value obtained within cRPA~\cite{hampel:2019,Seth_Georges:2017}. Different data-sets in Fig.~\ref{fig:lno_dc_comparison} correspond to DFT+DMFT calculations with different treatments of the DC correction, both OS and CSC, which we discuss in the following. 

We first focus on the data-set labeled ``CSC $n_\alpha^\text{DMFT}$'' (shown in red), which corresponds to the CSC calculation where the occupations entering the DC correction are calculated from the impurity occupations, and are updated in each DMFT iteration.
As discussed in Sec.~\ref{par:DC}, this is the correct way to perform such CSC DFT+DMFT calculations, since the converged $n_\alpha^\text{DMFT}$ give rise to the corrected charge density from which the KS potential is constructed within the DFT step. 
It can be seen, that the transition to the CDI occurs at $J=0.2$~eV, indicated by clear jumps in $\nu$ and $\bar{A}(0)$. The jump in $\nu$ is related to a drastic change in the DC potential difference between the Ni sites, since, for not too large $J$ (see also below), the DC correction tends to increase the charge disproportionation by further lowering the $e_g$ states on the more occupied LB site compared to the less occupied SB site. This is discussed and analyzed in more detail in Appendix~\ref{appendix}.

For further increasing $J$, $\nu$ stays almost constant until $J \approx 0.8$~eV, where $\nu$ decreases again. 
Finally, at around $J=1.2$~eV, the system becomes metallic again. 
This can be explained by the fact that for increasing $J$, the DC potential, proportional to $\bar{U}=U-\tfrac{5}{3}J$ (see Eq.~\eqref{eq:barU} for $M=2$), decreases, and eventually changes sign for $J=1.11$~eV where $\bar{U}=0$. 
Thus, above $J=1.11$\,eV the DC correction opposes the charge disproportionation by lowering the $e_g$ levels of the SB sites relative to the LB sites. 

Next we compare the CSC calculations with the simpler OS calculations. As discussed in Sec.~\ref{par:DC}, it is ambiguous whether to use the DMFT impurity occupations or the occupations of the Wannier functions obtained within DFT, $n_\alpha^\text{DFT}$, to evaluate the DC correction.
We first compare to the OS calculations where $n_\alpha^\text{DMFT}$ has been used for the DC correction (shown in green). 
It can be observed that in these OS calculations the system is already in the CDI state even for $J=0.2$~eV. In addition, a small shift to larger $\nu$ can be observed compared to the CSC case.
Thus, the tendency towards the CDI state is slightly stronger than in the CSC calculations.

In contrast, the OS calculations using $n_\alpha^\text{DFT}$ (shown in orange) leads to a significantly reduced $\nu$, which increases slowly with increasing $J$.
Moreover, for small $J<0.5$~eV, clear metallic behavior is observed, while from $J=0.5$ to 1.0~eV, the system undergoes the MIT, where eventually at $J=1.0$~eV the system is completely in the CDI state with $\nu > 1.0$. The occupations obtained in the initial DFT step are $n^\text{DFT}_\text{LB}\approx 1.15$ and $n^\text{DFT}_\text{SB}\approx 0.85$.

For comparison, we also perform CSC calculations where the DFT occupations are used for the DC correction (shown in purple). However, one should note, that these calculations are somewhat artificial, since the DFT Wannier orbital occupations loose their physical meaning in a CSC calculation, and are used here just to allow for a more systematic comparison between OS and CSC calculations.
One can see that overall the results of these calculations show similar behavior than the corresponding OS calculation using $n_\alpha^\text{DFT}$, albeit with a small further reduction of $\nu$.

The fixed structure calculations for LuNiO$_3$, show that performing CSC calculations leads to a small reduction of the charge disproportionation compared to OS calculations, if in both calculations the DMFT impurity occupations are used to determine the DC potential. Moreover, we find that the DC has a very strong effect, so that a OS calculation with DFT occupations significantly underestimates the tendency towards charge disproportionation compared to the ``correct'' CSC calculation.

Overall, we conclude that CSC has a small, but certainly not negligible influence on the DFT+DMFT calculations for LuNiO$_3$, reducing $\nu$ by approx. 10\%.
However, this only holds if DMFT occupations are used in the OS calculation to evaluate the DC correction.
If DFT occupations are used in the OS calculation, then the tendency towards the CDI state is significantly weakened, indicated by the much smaller $\nu$, which is related to the smaller difference in the DC potential shift.
However, compared to a hypothetical CSC calculation also using $n_\alpha^\text{DFT}$ for the DC correction, $\nu$ is again slightly enhanced in the OS calculation.
Thus, one can clearly distinguish between the effect of the DC correction, and the effect of the charge density correction in the CSC calculation. 
The latter tends to reduce the charge disproportionation, independently of the chosen DC scheme, and analogous to reducing the orbital polarization in the case of \cvo{} discussed in Sec.~\ref{sec:cvo}.
Finally, our results also indicate that the OS calculations using DMFT occupations for the DC correction already provide a good approximation for the CSC calculation, even though they slightly overestimate the SB/LB splitting and thus the tendency towards the CDI state.

\subsubsection{Influence on energetics}

\begin{figure}[t]
    \centering
    \includegraphics[width=1.0\linewidth]{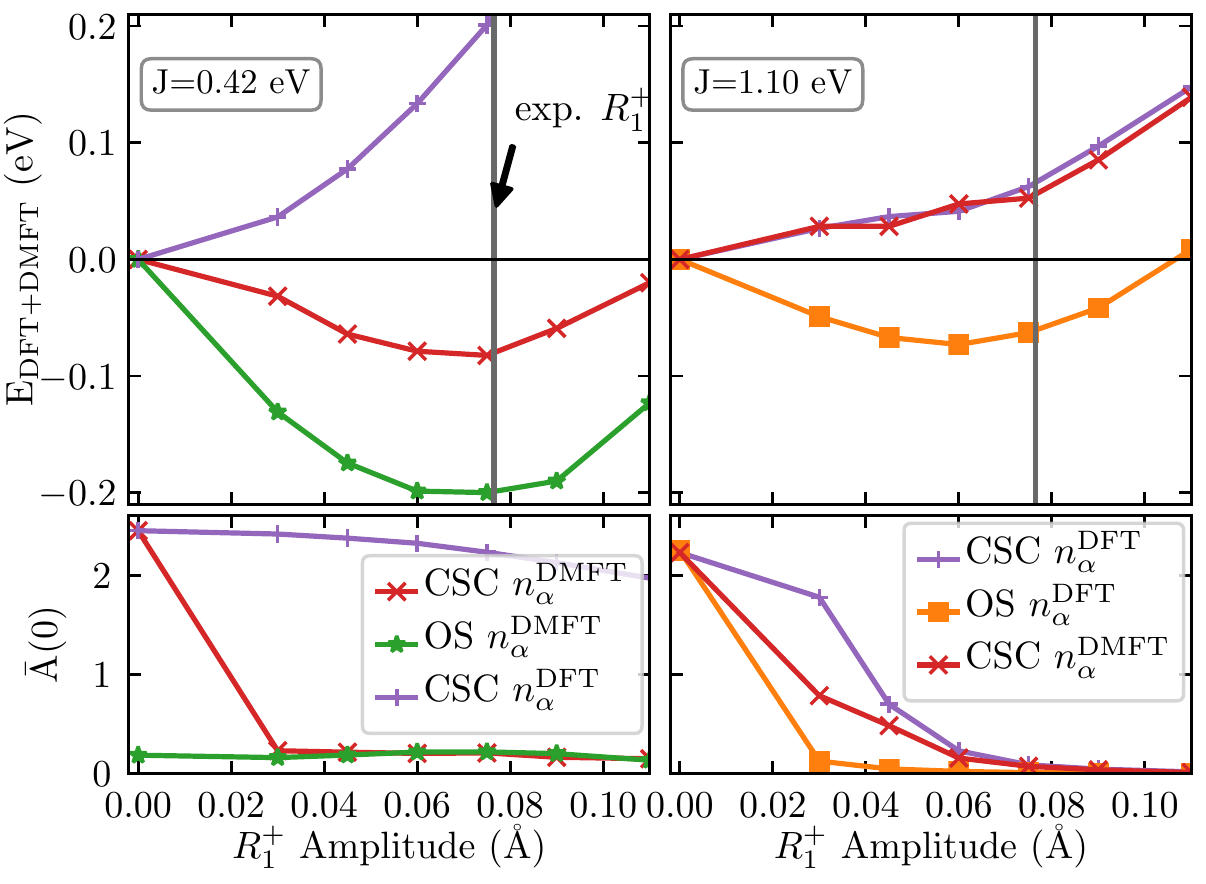}
    \caption{Comparison of energetics from DFT+DMFT for LuNiO$_3$ as function of the $R_1^+$ amplitude. Calculations without CSC are labeled OS, both in combination with DC occupations obtained from DFT ($n_\alpha^{\text{DFT}}$) or with occupations obtained from each DMFT step ($n_\alpha^{\text{DMFT}}$). The left panels show results for small $J=0.42$~eV and the right panels for large $J=1.1$~eV, where the upper panel shows the energy as function of the $R_1^+$ amplitude and the panels at the bottom the corresponding spectral weight at the Fermi level.}
    \label{fig:lno_energetics}
\end{figure}

Another important aspect is the influence of charge self-consistency in total energy calculations for different $R_1^+$ amplitudes, i.e., for different amplitudes of the structural breathing mode distortion. As the $R_1^+$ amplitude, and thus $\nu$, changes, the DC potential and energy correction change accordingly. 
In addition, within the CSC calculation, the Hartree energy and other DFT energy contributions are evaluated from the corrected, self-consistent charge density. Strictly speaking, only a full CSC calculation gives physical meaningful total energies~\cite{Kotliar:2006}, but nevertheless we discuss the difference here to better understand the influence of performing full CSC calculation~\cite{Park2014short,PhysRevB.76.235101}.
To analyze the resulting effects, we again use $U=1.85$~eV and two different values for $J$, 0.42~eV (the cRPA value) and 1.1~eV (where $\bar{U}\approx 0$ and thus the DC correction vanishes). For both cases, we compare OS with CSC calculations with different treatments of the DC correction, as introduced above.
The results are shown in Fig.~\ref{fig:lno_energetics}, where the top panels show the total energy as function of the $R_1^+$ amplitude, and the bottom panels show the corresponding $\bar{A}(0)$.

For the smaller value, $J=0.42$\,eV, both the OS (green) and CSC (red) result in an energy minimum at a finite $R_1^+$ amplitude close to the experimental value (indicated by the vertical line). However, the OS calculation exhibits a much stronger response on the $R_1^+$ amplitude, and hence shows a significantly deeper energy minimum. In contrast, the ``artificial'' CSC calculation using $n_\alpha^{\text{DFT}}$ for the DC correction (purple), exhibits no energy minimum for $R_1^+>0$.
Furthermore, the ``correct'' CSC calculation using $n_\alpha^\text{DMFT}$ undergoes a MIT to the CDI between $R_1^+=0$ and $R_1^+=0.03$\,\AA, while the corresponding OS calculation is already insulating without structural distortion and the CSC calculation with $n_\alpha^\text{DFT}$ remains metallic for any calculated $R_1^+$ amplitude.

For $J=1.1$~eV, both CSC calculations, done either with DFT (purple) or DMFT occupations (red), agree very well (due to $\bar{U} \approx 0$ in the DC) and do not result in a stable finite breathing mode amplitude, even though both undergo a MIT at around $R_1^+=0.03$~\r{A} and exhibit a large charge disproportionation $\nu$ in the insulating state.
In contrast, the OS calculation (orange), shows a stronger response, and predicts a breathing mode amplitude of $R_1^+=0.06$~\r{A}. 
Note that here we used $n_\alpha^\text{DFT}$ for the DC correction, but the same result would be obtained using $n_\alpha^\text{DMFT}$, due to $\bar{U}\approx 0$.

These results show that, even though the effect of charge self-consistency on $\nu$ for fixed crystal structure seems to be relatively minor, the effect on the energetics can be quite drastic, such that one can obtain a finite breathing mode distortion within a OS calculation, while the CSC calculation does not exhibit an energy minimum for $R_1^+>0$. 

\section{Summary}
\label{sec:summary}

We have studied the effect of charge self-consistency and the role of the DC correction within CSC DFT+DMFT calculations in two representative examples of transition metal oxides, using only a minimal correlated subspace corresponding to few frontier bands around the Fermi level. 
Our goal is to better understand in which cases charge self-consistency is really required in order to obtain accurate results, and in which cases a computationally much cheaper OS calculation might be sufficient.  

For \cvo{}, we find that the strong orbital polarization in the insulating phase under tensile strain is significantly overestimated by about 30\,\% in OS compared to CSC calculations, in agreement with similar calculations for SrVO$_3$ in Refs.~\onlinecite{Bhandary:2016,Schuler_Aichhorn:2018}.
This has a small but noticeable effect on $U_{\text{MIT}}$, the critical $U$ for the MIT, which is slightly underestimated in the OS calculations.
In contrast, for the unstrained system, where the orbital polarization is much smaller, the difference between CSC and OS calculations is nearly negligible, even though also in this case the orbital polarization is slightly overestimated in OS calculations. 
Furthermore, we also compared OS calculations using PLO-based and MLWF-based schemes for constructing the correlated subspace, and found very good agreement between the two methods.

While for \cvo{} all TM sites are symmetry-equivalent, and thus the site-dependent but orbitally-independent DC correction does not affect the results, for the second example investigated in this work, LuNiO$_3$, the DC correction becomes rather important. Here, we find that if DMFT occupations are used to evaluate the DC correction in the OS calculation, one can obtain results that are in rather good agreement with the CSC calculation, even though the charge disproportionation $\nu$ is overestimated by $~\sim 10$\,\%. Thus, similar to reducing the orbital polarization for strained \cvo{}, including charge self-consistency leads to a somewhat more homogeneous charge distribution compared to a OS calculation. 
Nevertheless, it appears that in order to obtain qualitative insights or general trends, OS calculations can be a reasonable approximation, even in charge ordered systems, if the DMFT occupations are used for the DC. 
However, our analysis of the energetics of the breathing mode distortion shows that for certain observables, such as the total energy and resulting structural distortions, charge self-consistency can be crucial. 
For example in the case of LuNiO$_3$, OS calculations overestimate the response on the $R_1^+$ mode, in the most extreme case leading to a stable finite breathing mode amplitude, which is absent in the CSC calculation.
In this case it is is inevitable to perform a full CSC calculation to obtain reliable results.

In summary, the effect of charge self-consistency is mainly to reduce a potential site or orbital polarization by favoring a more ``homogeneous'' distribution of electrons over all sites and/or orbitals. For the cases studied in this work, this results in a weak to moderate charge redistribution, which can be quantitatively relevant, depending on the specific application. In particular for total energy calculations, which depend on a subtle balance between different contributions, charge self-consistency can be crucial to obtain quantitatively and even qualitatively correct results. Nevertheless, it appears that cheaper OS calculations are often sufficient to gain insight into the system properties on a qualitative level, even though the, in principle ambiguous, choice of DFT or DMFT occupations to evaluate the DC correction in the OS calculations can become crucial. In the present examples, the use of DMFT occupations provided better agreement with the full CSC calculation, but in other cases this approach might also severely overestimate the electron transfer between inequivalent sites.

We hope that our detailed analysis of two specifically selected cases, provides useful insights for future DFT+DMFT studies of related material systems, thus allowing the treatment of larger and more complex materials systems by avoiding the higher computational cost of a CSC calculation when possible.

\begin{acknowledgments}
This work was supported by ETH Zurich and the Swiss National Science Foundation through NCCR-MARVEL. Calculations have been performed on the cluster ``Piz Daint'' hosted by the Swiss National Supercomputing Centre.
\end{acknowledgments}

\appendix

\section{Influence of the DC on the effective inter-site splitting}
\label{appendix}
In this appendix we explicitly show, how the DC corrections affects the $e_g$ level splitting between the two inequivalent Ni sites in the charge disproportionated state, which in turn controls the tendency to form a CDI state in the rare earth nickelates.

As outlined in the main text, \citet{Subedi:2015en} found that the CDI state emerges in the frontier $e_g$ model for nickelates when the following inequality is satisfied (derived from the the atomic limit):
\begin{align}
U - 3J \lesssim \Delta_s \quad .
\label{eq:neg-charge-transfer}
\end{align}
Here, $\Delta_s$ is the ``bare'' site splitting, i.e., the difference in the average $e_g$ orbital energy between the SB and LB Ni sites, and is given as:
\begin{align}
\Delta_s = \Delta_s^{\text{DFT}} - \Delta_s^{\text{DC}} \quad ,
\label{eq:deltas}
\end{align}
where the first term, $\Delta_s^{\text{DFT}}$, denotes the corresponding splitting obtained within DFT from the on-site energies of the Wannier functions, and is found to be $\approx 0.25$~eV for $R_1^+=0.075$~\r{A} in LuNiO$_3$. The second term, $\Delta_s^{\text{DC}}$, arises from the difference in the DC potential between the SB and LB sites:
\begin{align}
\Delta_s^{\text{DC}} = \Sigma_{\text{dc,SB}} - \Sigma_{\text{dc,LB}} \quad .
\label{eq:deltas_dc}
\end{align}

\begin{figure}[t]
    \centering
    \includegraphics[width=0.8\linewidth]{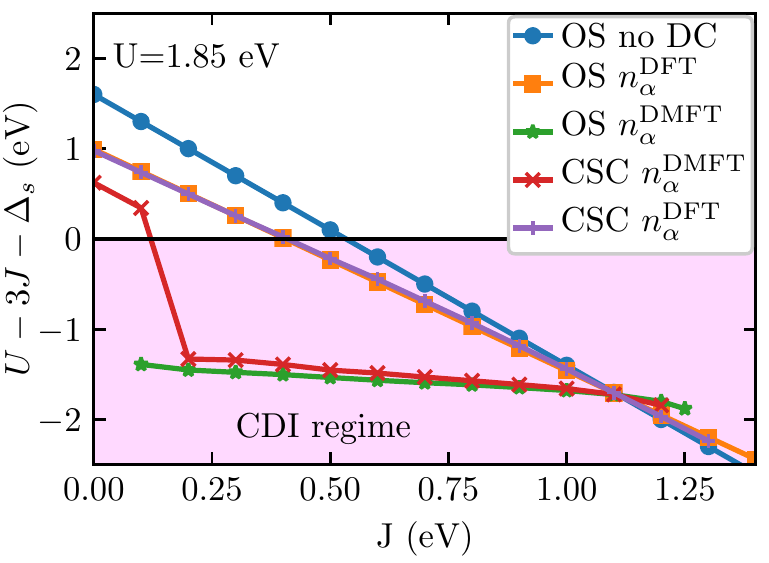}
    \caption{$U-3J-\Delta_s$ as function of $J$ for LuNiO$_3$ with the experimental $R_1^+$ amplitude~\cite{Alonso:2001bs}, corresponding to the calculations shown in Fig.~\ref{fig:lno_dc_comparison}. $U-3J-\Delta_s$ is shown for the different flavors of DC ($n_\alpha^{\text{DFT}}$ vs. $n_\alpha^{\text{DMFT}}$) and for OS and CSC calculations. If $U-3J-\Delta_s<0$ the CDI state is favored (magenta shaded area).}
    \label{fig:lno_u3j_ds}
\end{figure}

To further elucidate the interplay of the site-dependent DC potential in our DFT+DMFT calculations for LuNiO$_3$, we analyzed the behavior of the critical quantity $U - 3J - \Delta_s$ for obtaining a CDI state as function of $J$ for the different DC schemes applied in our study. 
The behavior of $U-3J-\Delta_s$ is depicted in Fig.~\ref{fig:lno_u3j_ds} for the different flavors of DC, and for OS and CSC calculations at fixed $U=1.85$~eV and $\Delta_s^{\text{DFT}}=0.25$~eV. The parameter regime which corresponds to the CDI in the atomic limit is highlighted in magenta. To directly compare our calculations with Ref.~\onlinecite{Subedi:2015en} we also performed OS calculations without applying any DC correction.

For the OS calculation without DC, $\Delta_s^{\text{DC}}=0$ (blue circles), $U-3J-\Delta_s$ becomes negative for $J \leq 0.53$~eV. Then, for the calculations with DC evaluated with DFT occupations (OS: orange squares, CSC: purple crosses) this happens at a slightly smaller $J$ value, leading to an increased response in $\nu(J)$ compared to the calculations without DC correction.
This shows, that the DC correction enhances the CDI state for positive values of $\bar{U}$ ($J<1.11$\,eV), since then $\Delta_s^\text{DC}$ is negative thus $\Delta_s > \Delta_s^\text{DFT}$.

The strong tendency to form the CDI state in the calculations with $n_\alpha^{\text{DMFT}}$ can be explained along the same lines. It can be seen that in the CSC calculations (red crosses), the CDI regime is entered already for $J=0.2$~eV, and in the OS calculations (green stars) for even smaller $J$. Importantly, it can be seen that in these cases the DC potential jumps at the MIT, strongly favoring the CDI state. We note that of course the underlying atomic limit consideration neglects the important effect of the bandwidth, but it nevertheless can give a transparent pciture of the underlying physics.

\bibliography{bibfile}

\begin{thebibliography}{73}%
\makeatletter
\providecommand \@ifxundefined [1]{%
 \@ifx{#1\undefined}
}%
\providecommand \@ifnum [1]{%
 \ifnum #1\expandafter \@firstoftwo
 \else \expandafter \@secondoftwo
 \fi
}%
\providecommand \@ifx [1]{%
 \ifx #1\expandafter \@firstoftwo
 \else \expandafter \@secondoftwo
 \fi
}%
\providecommand \natexlab [1]{#1}%
\providecommand \enquote  [1]{``#1''}%
\providecommand \bibnamefont  [1]{#1}%
\providecommand \bibfnamefont [1]{#1}%
\providecommand \citenamefont [1]{#1}%
\providecommand \href@noop [0]{\@secondoftwo}%
\providecommand \href [0]{\begingroup \@sanitize@url \@href}%
\providecommand \@href[1]{\@@startlink{#1}\@@href}%
\providecommand \@@href[1]{\endgroup#1\@@endlink}%
\providecommand \@sanitize@url [0]{\catcode `\\12\catcode `\$12\catcode
  `\&12\catcode `\#12\catcode `\^12\catcode `\_12\catcode `\%12\relax}%
\providecommand \@@startlink[1]{}%
\providecommand \@@endlink[0]{}%
\providecommand \url  [0]{\begingroup\@sanitize@url \@url }%
\providecommand \@url [1]{\endgroup\@href {#1}{\urlprefix }}%
\providecommand \urlprefix  [0]{URL }%
\providecommand \Eprint [0]{\href }%
\providecommand \doibase [0]{http://dx.doi.org/}%
\providecommand \selectlanguage [0]{\@gobble}%
\providecommand \bibinfo  [0]{\@secondoftwo}%
\providecommand \bibfield  [0]{\@secondoftwo}%
\providecommand \translation [1]{[#1]}%
\providecommand \BibitemOpen [0]{}%
\providecommand \bibitemStop [0]{}%
\providecommand \bibitemNoStop [0]{.\EOS\space}%
\providecommand \EOS [0]{\spacefactor3000\relax}%
\providecommand \BibitemShut  [1]{\csname bibitem#1\endcsname}%
\let\auto@bib@innerbib\@empty
\bibitem [{\citenamefont {Held}(2007)}]{held2007}%
  \BibitemOpen
  \bibfield  {author} {\bibinfo {author} {\bibfnamefont {K.}~\bibnamefont
  {Held}},\ }\href {\doibase 10.1080/00018730701619647} {\bibfield  {journal}
  {\bibinfo  {journal} {Advances in Physics}\ }\textbf {\bibinfo {volume}
  {56}},\ \bibinfo {pages} {829} (\bibinfo {year} {2007})}\BibitemShut
  {NoStop}%
\bibitem [{\citenamefont {Grieger}\ \emph {et~al.}(2012)\citenamefont
  {Grieger}, \citenamefont {Piefke}, \citenamefont {Peil},\ and\ \citenamefont
  {Lechermann}}]{Grieger_et_al:2012}%
  \BibitemOpen
  \bibfield  {author} {\bibinfo {author} {\bibfnamefont {D.}~\bibnamefont
  {Grieger}}, \bibinfo {author} {\bibfnamefont {C.}~\bibnamefont {Piefke}},
  \bibinfo {author} {\bibfnamefont {O.~E.}\ \bibnamefont {Peil}}, \ and\
  \bibinfo {author} {\bibfnamefont {F.}~\bibnamefont {Lechermann}},\ }\href
  {\doibase 10.1103/PhysRevB.86.155121} {\bibfield  {journal} {\bibinfo
  {journal} {Physical Review B}\ }\textbf {\bibinfo {volume} {86}},\ \bibinfo
  {pages} {155121} (\bibinfo {year} {2012})}\BibitemShut {NoStop}%
\bibitem [{\citenamefont {Adler}\ \emph {et~al.}(2019)\citenamefont {Adler},
  \citenamefont {Kang}, \citenamefont {Yee},\ and\ \citenamefont
  {Kotliar}}]{Adler_et_al:2019}%
  \BibitemOpen
  \bibfield  {author} {\bibinfo {author} {\bibfnamefont {R.}~\bibnamefont
  {Adler}}, \bibinfo {author} {\bibfnamefont {C.-J.}\ \bibnamefont {Kang}},
  \bibinfo {author} {\bibfnamefont {C.-H.}\ \bibnamefont {Yee}}, \ and\
  \bibinfo {author} {\bibfnamefont {G.}~\bibnamefont {Kotliar}},\ }\href
  {\doibase https://doi.org/10.1088/1361-6633/aadca4} {\bibfield  {journal}
  {\bibinfo  {journal} {Reports on Progress in Physics}\ }\textbf {\bibinfo
  {volume} {82}},\ \bibinfo {pages} {012504} (\bibinfo {year}
  {2019})}\BibitemShut {NoStop}%
\bibitem [{\citenamefont {Ishida}\ and\ \citenamefont
  {Liebsch}(2009)}]{Ishida:2009}%
  \BibitemOpen
  \bibfield  {author} {\bibinfo {author} {\bibfnamefont {H.}~\bibnamefont
  {Ishida}}\ and\ \bibinfo {author} {\bibfnamefont {A.}~\bibnamefont
  {Liebsch}},\ }\href {\doibase 10.1103/PhysRevB.79.045130} {\bibfield
  {journal} {\bibinfo  {journal} {Physical Review B}\ }\textbf {\bibinfo
  {volume} {79}},\ \bibinfo {pages} {045130} (\bibinfo {year}
  {2009})}\BibitemShut {NoStop}%
\bibitem [{\citenamefont {Zhong}\ \emph {et~al.}(2015)\citenamefont {Zhong},
  \citenamefont {Wallerberger}, \citenamefont {Tomczak}, \citenamefont
  {Taranto}, \citenamefont {Parragh}, \citenamefont {Toschi}, \citenamefont
  {Sangiovanni},\ and\ \citenamefont {Held}}]{Zhong:2015}%
  \BibitemOpen
  \bibfield  {author} {\bibinfo {author} {\bibfnamefont {Z.}~\bibnamefont
  {Zhong}}, \bibinfo {author} {\bibfnamefont {M.}~\bibnamefont {Wallerberger}},
  \bibinfo {author} {\bibfnamefont {J.~M.}\ \bibnamefont {Tomczak}}, \bibinfo
  {author} {\bibfnamefont {C.}~\bibnamefont {Taranto}}, \bibinfo {author}
  {\bibfnamefont {N.}~\bibnamefont {Parragh}}, \bibinfo {author} {\bibfnamefont
  {A.}~\bibnamefont {Toschi}}, \bibinfo {author} {\bibfnamefont
  {G.}~\bibnamefont {Sangiovanni}}, \ and\ \bibinfo {author} {\bibfnamefont
  {K.}~\bibnamefont {Held}},\ }\href {\doibase 10.1103/PhysRevLett.114.246401}
  {\bibfield  {journal} {\bibinfo  {journal} {Physical Review Letters}\
  }\textbf {\bibinfo {volume} {114}},\ \bibinfo {pages} {246401} (\bibinfo
  {year} {2015})}\BibitemShut {NoStop}%
\bibitem [{\citenamefont {Lechermann}(2018)}]{Lechermann:2018}%
  \BibitemOpen
  \bibfield  {author} {\bibinfo {author} {\bibfnamefont {F.}~\bibnamefont
  {Lechermann}},\ }in\ \href {\doibase 10.1007/978-3-319-50257-1_80-1} {\emph
  {\bibinfo {booktitle} {Handbook of Materials Modeling}}}\ (\bibinfo
  {publisher} {Springer International Publishing},\ \bibinfo {year} {2018})\
  pp.\ \bibinfo {pages} {1--20}\BibitemShut {NoStop}%
\bibitem [{\citenamefont {Beck}\ and\ \citenamefont
  {Ederer}(2019)}]{Beck:2019}%
  \BibitemOpen
  \bibfield  {author} {\bibinfo {author} {\bibfnamefont {S.}~\bibnamefont
  {Beck}}\ and\ \bibinfo {author} {\bibfnamefont {C.}~\bibnamefont {Ederer}},\
  }\href {\doibase 10.1103/PhysRevMaterials.3.095001} {\bibfield  {journal}
  {\bibinfo  {journal} {Phys. Rev. Materials}\ }\textbf {\bibinfo {volume}
  {3}},\ \bibinfo {pages} {095001} (\bibinfo {year} {2019})}\BibitemShut
  {NoStop}%
\bibitem [{\citenamefont {Backes}\ \emph {et~al.}(2016)\citenamefont {Backes},
  \citenamefont {R\"odel}, \citenamefont {Fortuna}, \citenamefont
  {Frantzeskakis}, \citenamefont {Le~F\`evre}, \citenamefont {Bertran},
  \citenamefont {Kobayashi}, \citenamefont {Yukawa}, \citenamefont
  {Mitsuhashi}, \citenamefont {Kitamura}, \citenamefont {Horiba}, \citenamefont
  {Kumigashira}, \citenamefont {Saint-Martin}, \citenamefont {Fouchet},
  \citenamefont {Berini}, \citenamefont {Dumont}, \citenamefont {Kim},
  \citenamefont {Lechermann}, \citenamefont {Jeschke}, \citenamefont
  {Rozenberg}, \citenamefont {Valent\'{\i}},\ and\ \citenamefont
  {Santander-Syro}}]{Backes_et_al:2016}%
  \BibitemOpen
  \bibfield  {author} {\bibinfo {author} {\bibfnamefont {S.}~\bibnamefont
  {Backes}}, \bibinfo {author} {\bibfnamefont {T.~C.}\ \bibnamefont {R\"odel}},
  \bibinfo {author} {\bibfnamefont {F.}~\bibnamefont {Fortuna}}, \bibinfo
  {author} {\bibfnamefont {E.}~\bibnamefont {Frantzeskakis}}, \bibinfo {author}
  {\bibfnamefont {P.}~\bibnamefont {Le~F\`evre}}, \bibinfo {author}
  {\bibfnamefont {F.}~\bibnamefont {Bertran}}, \bibinfo {author} {\bibfnamefont
  {M.}~\bibnamefont {Kobayashi}}, \bibinfo {author} {\bibfnamefont
  {R.}~\bibnamefont {Yukawa}}, \bibinfo {author} {\bibfnamefont
  {T.}~\bibnamefont {Mitsuhashi}}, \bibinfo {author} {\bibfnamefont
  {M.}~\bibnamefont {Kitamura}}, \bibinfo {author} {\bibfnamefont
  {K.}~\bibnamefont {Horiba}}, \bibinfo {author} {\bibfnamefont
  {H.}~\bibnamefont {Kumigashira}}, \bibinfo {author} {\bibfnamefont
  {R.}~\bibnamefont {Saint-Martin}}, \bibinfo {author} {\bibfnamefont
  {A.}~\bibnamefont {Fouchet}}, \bibinfo {author} {\bibfnamefont
  {B.}~\bibnamefont {Berini}}, \bibinfo {author} {\bibfnamefont
  {Y.}~\bibnamefont {Dumont}}, \bibinfo {author} {\bibfnamefont {A.~J.}\
  \bibnamefont {Kim}}, \bibinfo {author} {\bibfnamefont {F.}~\bibnamefont
  {Lechermann}}, \bibinfo {author} {\bibfnamefont {H.~O.}\ \bibnamefont
  {Jeschke}}, \bibinfo {author} {\bibfnamefont {M.~J.}\ \bibnamefont
  {Rozenberg}}, \bibinfo {author} {\bibfnamefont {R.}~\bibnamefont
  {Valent\'{\i}}}, \ and\ \bibinfo {author} {\bibfnamefont {A.~F.}\
  \bibnamefont {Santander-Syro}},\ }\href {\doibase 10.1103/PhysRevB.94.241110}
  {\bibfield  {journal} {\bibinfo  {journal} {Physical Review B}\ }\textbf
  {\bibinfo {volume} {94}},\ \bibinfo {pages} {241110} (\bibinfo {year}
  {2016})}\BibitemShut {NoStop}%
\bibitem [{\citenamefont {Sing}\ \emph {et~al.}(2017)\citenamefont {Sing},
  \citenamefont {Jeschke}, \citenamefont {Lechermann}, \citenamefont
  {Valent{\'\i}},\ and\ \citenamefont {Claessen}}]{Sing:2017}%
  \BibitemOpen
  \bibfield  {author} {\bibinfo {author} {\bibfnamefont {M.}~\bibnamefont
  {Sing}}, \bibinfo {author} {\bibfnamefont {H.~O.}\ \bibnamefont {Jeschke}},
  \bibinfo {author} {\bibfnamefont {F.}~\bibnamefont {Lechermann}}, \bibinfo
  {author} {\bibfnamefont {R.}~\bibnamefont {Valent{\'\i}}}, \ and\ \bibinfo
  {author} {\bibfnamefont {R.}~\bibnamefont {Claessen}},\ }\href@noop {}
  {\bibfield  {journal} {\bibinfo  {journal} {The European Physical Journal
  Special Topics}\ }\textbf {\bibinfo {volume} {226}},\ \bibinfo {pages} {2457}
  (\bibinfo {year} {2017})}\BibitemShut {NoStop}%
\bibitem [{\citenamefont {Souto-Casares}\ \emph {et~al.}(2019)\citenamefont
  {Souto-Casares}, \citenamefont {Spaldin},\ and\ \citenamefont
  {Ederer}}]{Souto-Casares:2019}%
  \BibitemOpen
  \bibfield  {author} {\bibinfo {author} {\bibfnamefont {J.}~\bibnamefont
  {Souto-Casares}}, \bibinfo {author} {\bibfnamefont {N.~A.}\ \bibnamefont
  {Spaldin}}, \ and\ \bibinfo {author} {\bibfnamefont {C.}~\bibnamefont
  {Ederer}},\ }\href {\doibase 10.1103/PhysRevB.100.085146} {\bibfield
  {journal} {\bibinfo  {journal} {Phys. Rev. B}\ }\textbf {\bibinfo {volume}
  {100}},\ \bibinfo {pages} {085146} (\bibinfo {year} {2019})}\BibitemShut
  {NoStop}%
\bibitem [{\citenamefont {Jacob}\ \emph {et~al.}(2010)\citenamefont {Jacob},
  \citenamefont {Haule},\ and\ \citenamefont {Kotliar}}]{Jacob:2010}%
  \BibitemOpen
  \bibfield  {author} {\bibinfo {author} {\bibfnamefont {D.}~\bibnamefont
  {Jacob}}, \bibinfo {author} {\bibfnamefont {K.}~\bibnamefont {Haule}}, \ and\
  \bibinfo {author} {\bibfnamefont {G.}~\bibnamefont {Kotliar}},\ }\href
  {\doibase 10.1103/PhysRevB.82.195115} {\bibfield  {journal} {\bibinfo
  {journal} {Physical Review B}\ }\textbf {\bibinfo {volume} {82}},\ \bibinfo
  {pages} {195115} (\bibinfo {year} {2010})}\BibitemShut {NoStop}%
\bibitem [{\citenamefont {Turkowski}\ \emph {et~al.}(2012)\citenamefont
  {Turkowski}, \citenamefont {Kabir}, \citenamefont {Nayyar},\ and\
  \citenamefont {Rahman}}]{Turkowski:2012}%
  \BibitemOpen
  \bibfield  {author} {\bibinfo {author} {\bibfnamefont {V.}~\bibnamefont
  {Turkowski}}, \bibinfo {author} {\bibfnamefont {A.}~\bibnamefont {Kabir}},
  \bibinfo {author} {\bibfnamefont {N.}~\bibnamefont {Nayyar}}, \ and\ \bibinfo
  {author} {\bibfnamefont {T.~S.}\ \bibnamefont {Rahman}},\ }\href {\doibase
  10.1063/1.3692613} {\bibfield  {journal} {\bibinfo  {journal} {The Journal of
  Chemical Physics}\ }\textbf {\bibinfo {volume} {136}},\ \bibinfo {pages}
  {114108} (\bibinfo {year} {2012})}\BibitemShut {NoStop}%
\bibitem [{\citenamefont {Bhandary}\ \emph {et~al.}(2016)\citenamefont
  {Bhandary}, \citenamefont {Assmann}, \citenamefont {Aichhorn},\ and\
  \citenamefont {Held}}]{Bhandary:2016}%
  \BibitemOpen
  \bibfield  {author} {\bibinfo {author} {\bibfnamefont {S.}~\bibnamefont
  {Bhandary}}, \bibinfo {author} {\bibfnamefont {E.}~\bibnamefont {Assmann}},
  \bibinfo {author} {\bibfnamefont {M.}~\bibnamefont {Aichhorn}}, \ and\
  \bibinfo {author} {\bibfnamefont {K.}~\bibnamefont {Held}},\ }\href {\doibase
  10.1103/PhysRevB.94.155131} {\bibfield  {journal} {\bibinfo  {journal}
  {Physical Review B}\ }\textbf {\bibinfo {volume} {94}},\ \bibinfo {pages}
  {155131} (\bibinfo {year} {2016})}\BibitemShut {NoStop}%
\bibitem [{\citenamefont {Sch{\"u}ler}\ \emph {et~al.}(2018)\citenamefont
  {Sch{\"u}ler}, \citenamefont {Peil}, \citenamefont {Kraberger}, \citenamefont
  {Pordzik}, \citenamefont {Marsman}, \citenamefont {Kresse}, \citenamefont
  {Wehling},\ and\ \citenamefont {Aichhorn}}]{Schuler_Aichhorn:2018}%
  \BibitemOpen
  \bibfield  {author} {\bibinfo {author} {\bibfnamefont {M.}~\bibnamefont
  {Sch{\"u}ler}}, \bibinfo {author} {\bibfnamefont {O.~E.}\ \bibnamefont
  {Peil}}, \bibinfo {author} {\bibfnamefont {G.~J.}\ \bibnamefont {Kraberger}},
  \bibinfo {author} {\bibfnamefont {R.}~\bibnamefont {Pordzik}}, \bibinfo
  {author} {\bibfnamefont {M.}~\bibnamefont {Marsman}}, \bibinfo {author}
  {\bibfnamefont {G.}~\bibnamefont {Kresse}}, \bibinfo {author} {\bibfnamefont
  {T.~O.}\ \bibnamefont {Wehling}}, \ and\ \bibinfo {author} {\bibfnamefont
  {M.}~\bibnamefont {Aichhorn}},\ }\href@noop {} {\bibfield  {journal}
  {\bibinfo  {journal} {Journal of Physics: Condensed Matter}\ }\textbf
  {\bibinfo {volume} {30}},\ \bibinfo {pages} {475901} (\bibinfo {year}
  {2018})}\BibitemShut {NoStop}%
\bibitem [{\citenamefont {Gu}\ \emph {et~al.}(2013)\citenamefont {Gu},
  \citenamefont {Laverock}, \citenamefont {Chen}, \citenamefont {Smith},
  \citenamefont {Wolf},\ and\ \citenamefont {Lu}}]{Gu_et_al:2013}%
  \BibitemOpen
  \bibfield  {author} {\bibinfo {author} {\bibfnamefont {M.}~\bibnamefont
  {Gu}}, \bibinfo {author} {\bibfnamefont {J.}~\bibnamefont {Laverock}},
  \bibinfo {author} {\bibfnamefont {B.}~\bibnamefont {Chen}}, \bibinfo {author}
  {\bibfnamefont {K.~E.}\ \bibnamefont {Smith}}, \bibinfo {author}
  {\bibfnamefont {S.~A.}\ \bibnamefont {Wolf}}, \ and\ \bibinfo {author}
  {\bibfnamefont {J.}~\bibnamefont {Lu}},\ }\href {\doibase 10.1063/1.4798963}
  {\bibfield  {journal} {\bibinfo  {journal} {Journal of Applied Physics}\
  }\textbf {\bibinfo {volume} {113}},\ \bibinfo {pages} {133704} (\bibinfo
  {year} {2013})}\BibitemShut {NoStop}%
\bibitem [{\citenamefont {Beck}\ \emph {et~al.}(2018)\citenamefont {Beck},
  \citenamefont {Sclauzero}, \citenamefont {Chopra},\ and\ \citenamefont
  {Ederer}}]{Beck:2018}%
  \BibitemOpen
  \bibfield  {author} {\bibinfo {author} {\bibfnamefont {S.}~\bibnamefont
  {Beck}}, \bibinfo {author} {\bibfnamefont {G.}~\bibnamefont {Sclauzero}},
  \bibinfo {author} {\bibfnamefont {U.}~\bibnamefont {Chopra}}, \ and\ \bibinfo
  {author} {\bibfnamefont {C.}~\bibnamefont {Ederer}},\ }\href {\doibase
  10.1103/PhysRevB.97.075107} {\bibfield  {journal} {\bibinfo  {journal}
  {Physical Review B}\ }\textbf {\bibinfo {volume} {97}},\ \bibinfo {pages}
  {075107} (\bibinfo {year} {2018})}\BibitemShut {NoStop}%
\bibitem [{\citenamefont {Catalano}\ \emph {et~al.}(2018)\citenamefont
  {Catalano}, \citenamefont {Gibert}, \citenamefont {Fowlie}, \citenamefont
  {{\'I}{\~n}iguez}, \citenamefont {Triscone},\ and\ \citenamefont
  {Kreisel}}]{catalano.review}%
  \BibitemOpen
  \bibfield  {author} {\bibinfo {author} {\bibfnamefont {S.}~\bibnamefont
  {Catalano}}, \bibinfo {author} {\bibfnamefont {M.}~\bibnamefont {Gibert}},
  \bibinfo {author} {\bibfnamefont {J.}~\bibnamefont {Fowlie}}, \bibinfo
  {author} {\bibfnamefont {J.}~\bibnamefont {{\'I}{\~n}iguez}}, \bibinfo
  {author} {\bibfnamefont {J.-M.}\ \bibnamefont {Triscone}}, \ and\ \bibinfo
  {author} {\bibfnamefont {J.}~\bibnamefont {Kreisel}},\ }\href
  {http://stacks.iop.org/0034-4885/81/i=4/a=046501} {\bibfield  {journal}
  {\bibinfo  {journal} {Reports on Progress in Physics}\ }\textbf {\bibinfo
  {volume} {81}},\ \bibinfo {pages} {046501} (\bibinfo {year}
  {2018})}\BibitemShut {NoStop}%
\bibitem [{\citenamefont {Wang}\ \emph {et~al.}(2012)\citenamefont {Wang},
  \citenamefont {Han}, \citenamefont {de' Medici}, \citenamefont {Park},
  \citenamefont {Marianetti},\ and\ \citenamefont {Millis}}]{Wang:2012}%
  \BibitemOpen
  \bibfield  {author} {\bibinfo {author} {\bibfnamefont {X.}~\bibnamefont
  {Wang}}, \bibinfo {author} {\bibfnamefont {M.~J.}\ \bibnamefont {Han}},
  \bibinfo {author} {\bibfnamefont {L.}~\bibnamefont {de' Medici}}, \bibinfo
  {author} {\bibfnamefont {H.}~\bibnamefont {Park}}, \bibinfo {author}
  {\bibfnamefont {C.~A.}\ \bibnamefont {Marianetti}}, \ and\ \bibinfo {author}
  {\bibfnamefont {A.~J.}\ \bibnamefont {Millis}},\ }\href {\doibase
  10.1103/PhysRevB.86.195136} {\bibfield  {journal} {\bibinfo  {journal}
  {Physical Review B}\ }\textbf {\bibinfo {volume} {86}},\ \bibinfo {pages}
  {195136} (\bibinfo {year} {2012})}\BibitemShut {NoStop}%
\bibitem [{\citenamefont {Dang}\ \emph {et~al.}(2014)\citenamefont {Dang},
  \citenamefont {Ai}, \citenamefont {Millis},\ and\ \citenamefont
  {Marianetti}}]{Dang:2014}%
  \BibitemOpen
  \bibfield  {author} {\bibinfo {author} {\bibfnamefont {H.~T.}\ \bibnamefont
  {Dang}}, \bibinfo {author} {\bibfnamefont {X.}~\bibnamefont {Ai}}, \bibinfo
  {author} {\bibfnamefont {A.~J.}\ \bibnamefont {Millis}}, \ and\ \bibinfo
  {author} {\bibfnamefont {C.~A.}\ \bibnamefont {Marianetti}},\ }\href
  {\doibase 10.1103/PhysRevB.90.125114} {\bibfield  {journal} {\bibinfo
  {journal} {Physical Review B}\ }\textbf {\bibinfo {volume} {90}},\ \bibinfo
  {pages} {125114} (\bibinfo {year} {2014})}\BibitemShut {NoStop}%
\bibitem [{\citenamefont {Park}\ \emph
  {et~al.}(2014{\natexlab{a}})\citenamefont {Park}, \citenamefont {Millis},\
  and\ \citenamefont {Marianetti}}]{Park:2014}%
  \BibitemOpen
  \bibfield  {author} {\bibinfo {author} {\bibfnamefont {H.}~\bibnamefont
  {Park}}, \bibinfo {author} {\bibfnamefont {A.~J.}\ \bibnamefont {Millis}}, \
  and\ \bibinfo {author} {\bibfnamefont {C.~A.}\ \bibnamefont {Marianetti}},\
  }\href {\doibase 10.1103/PhysRevB.90.235103} {\bibfield  {journal} {\bibinfo
  {journal} {Physical Review B}\ }\textbf {\bibinfo {volume} {90}},\ \bibinfo
  {pages} {235103} (\bibinfo {year} {2014}{\natexlab{a}})}\BibitemShut
  {NoStop}%
\bibitem [{\citenamefont {Aichhorn}\ \emph {et~al.}(2009)\citenamefont
  {Aichhorn}, \citenamefont {Pourovskii}, \citenamefont {Vildosola},
  \citenamefont {Ferrero}, \citenamefont {Parcollet}, \citenamefont {Miyake},
  \citenamefont {Georges},\ and\ \citenamefont {Biermann}}]{Aichhorn:2009}%
  \BibitemOpen
  \bibfield  {author} {\bibinfo {author} {\bibfnamefont {M.}~\bibnamefont
  {Aichhorn}}, \bibinfo {author} {\bibfnamefont {L.}~\bibnamefont
  {Pourovskii}}, \bibinfo {author} {\bibfnamefont {V.}~\bibnamefont
  {Vildosola}}, \bibinfo {author} {\bibfnamefont {M.}~\bibnamefont {Ferrero}},
  \bibinfo {author} {\bibfnamefont {O.}~\bibnamefont {Parcollet}}, \bibinfo
  {author} {\bibfnamefont {T.}~\bibnamefont {Miyake}}, \bibinfo {author}
  {\bibfnamefont {A.}~\bibnamefont {Georges}}, \ and\ \bibinfo {author}
  {\bibfnamefont {S.}~\bibnamefont {Biermann}},\ }\href {\doibase
  10.1103/PhysRevB.80.085101} {\bibfield  {journal} {\bibinfo  {journal}
  {Physical Review B}\ }\textbf {\bibinfo {volume} {80}},\ \bibinfo {pages}
  {085101} (\bibinfo {year} {2009})}\BibitemShut {NoStop}%
\bibitem [{\citenamefont {Aichhorn}\ \emph {et~al.}(2011)\citenamefont
  {Aichhorn}, \citenamefont {Pourovskii},\ and\ \citenamefont
  {Georges}}]{Aichhorn:2011}%
  \BibitemOpen
  \bibfield  {author} {\bibinfo {author} {\bibfnamefont {M.}~\bibnamefont
  {Aichhorn}}, \bibinfo {author} {\bibfnamefont {L.}~\bibnamefont
  {Pourovskii}}, \ and\ \bibinfo {author} {\bibfnamefont {A.}~\bibnamefont
  {Georges}},\ }\href {\doibase 10.1103/PhysRevB.84.054529} {\bibfield
  {journal} {\bibinfo  {journal} {Physical Review B}\ }\textbf {\bibinfo
  {volume} {84}},\ \bibinfo {pages} {054529} (\bibinfo {year}
  {2011})}\BibitemShut {NoStop}%
\bibitem [{\citenamefont {Karolak}\ \emph {et~al.}(2010)\citenamefont
  {Karolak}, \citenamefont {Ulm}, \citenamefont {Wehling}, \citenamefont
  {Mazurenko}, \citenamefont {Poteryaev},\ and\ \citenamefont
  {Lichtenstein}}]{Karolak:2010}%
  \BibitemOpen
  \bibfield  {author} {\bibinfo {author} {\bibfnamefont {M.}~\bibnamefont
  {Karolak}}, \bibinfo {author} {\bibfnamefont {G.}~\bibnamefont {Ulm}},
  \bibinfo {author} {\bibfnamefont {T.}~\bibnamefont {Wehling}}, \bibinfo
  {author} {\bibfnamefont {V.}~\bibnamefont {Mazurenko}}, \bibinfo {author}
  {\bibfnamefont {A.}~\bibnamefont {Poteryaev}}, \ and\ \bibinfo {author}
  {\bibfnamefont {A.}~\bibnamefont {Lichtenstein}},\ }\href {\doibase
  https://doi.org/10.1016/j.elspec.2010.05.021} {\bibfield  {journal} {\bibinfo
   {journal} {Journal of Electron Spectroscopy and Related Phenomena}\ }\textbf
  {\bibinfo {volume} {181}},\ \bibinfo {pages} {11 } (\bibinfo {year}
  {2010})},\ \bibinfo {note} {proceedings of International Workshop on Strong
  Correlations and Angle-Resolved Photoemission Spectroscopy 2009}\BibitemShut
  {NoStop}%
\bibitem [{\citenamefont {Lechermann}\ \emph {et~al.}(2006)\citenamefont
  {Lechermann}, \citenamefont {Georges}, \citenamefont {Poteryaev},
  \citenamefont {Biermann}, \citenamefont {Posternak}, \citenamefont
  {Yamasaki},\ and\ \citenamefont {Andersen}}]{PhysRevB.74.125120}%
  \BibitemOpen
  \bibfield  {author} {\bibinfo {author} {\bibfnamefont {F.}~\bibnamefont
  {Lechermann}}, \bibinfo {author} {\bibfnamefont {A.}~\bibnamefont {Georges}},
  \bibinfo {author} {\bibfnamefont {A.}~\bibnamefont {Poteryaev}}, \bibinfo
  {author} {\bibfnamefont {S.}~\bibnamefont {Biermann}}, \bibinfo {author}
  {\bibfnamefont {M.}~\bibnamefont {Posternak}}, \bibinfo {author}
  {\bibfnamefont {A.}~\bibnamefont {Yamasaki}}, \ and\ \bibinfo {author}
  {\bibfnamefont {O.~K.}\ \bibnamefont {Andersen}},\ }\href {\doibase
  10.1103/PhysRevB.74.125120} {\bibfield  {journal} {\bibinfo  {journal}
  {Physical Review B}\ }\textbf {\bibinfo {volume} {74}},\ \bibinfo {pages}
  {125120} (\bibinfo {year} {2006})}\BibitemShut {NoStop}%
\bibitem [{\citenamefont {Solovyev}\ and\ \citenamefont
  {Terakura}(1998)}]{Solovyev/Terakura:1998}%
  \BibitemOpen
  \bibfield  {author} {\bibinfo {author} {\bibfnamefont {I.~V.}\ \bibnamefont
  {Solovyev}}\ and\ \bibinfo {author} {\bibfnamefont {K.}~\bibnamefont
  {Terakura}},\ }\href {\doibase 10.1103/PhysRevB.58.15496} {\bibfield
  {journal} {\bibinfo  {journal} {Phys. Rev. B}\ }\textbf {\bibinfo {volume}
  {58}},\ \bibinfo {pages} {15496} (\bibinfo {year} {1998})}\BibitemShut
  {NoStop}%
\bibitem [{\citenamefont {Kotliar}\ \emph {et~al.}(2006)\citenamefont
  {Kotliar}, \citenamefont {Savrasov}, \citenamefont {Haule}, \citenamefont
  {Oudovenko}, \citenamefont {Parcollet},\ and\ \citenamefont
  {Marianetti}}]{Kotliar:2006}%
  \BibitemOpen
  \bibfield  {author} {\bibinfo {author} {\bibfnamefont {G.}~\bibnamefont
  {Kotliar}}, \bibinfo {author} {\bibfnamefont {S.~Y.}\ \bibnamefont
  {Savrasov}}, \bibinfo {author} {\bibfnamefont {K.}~\bibnamefont {Haule}},
  \bibinfo {author} {\bibfnamefont {V.~S.}\ \bibnamefont {Oudovenko}}, \bibinfo
  {author} {\bibfnamefont {O.}~\bibnamefont {Parcollet}}, \ and\ \bibinfo
  {author} {\bibfnamefont {C.~A.}\ \bibnamefont {Marianetti}},\ }\href
  {\doibase 10.1103/RevModPhys.78.865} {\bibfield  {journal} {\bibinfo
  {journal} {Reviews of Modern Physics}\ }\textbf {\bibinfo {volume} {78}},\
  \bibinfo {pages} {865} (\bibinfo {year} {2006})}\BibitemShut {NoStop}%
\bibitem [{\citenamefont {Haule}(2015)}]{Haule:2015_exactDC}%
  \BibitemOpen
  \bibfield  {author} {\bibinfo {author} {\bibfnamefont {K.}~\bibnamefont
  {Haule}},\ }\href {\doibase 10.1103/PhysRevLett.115.196403} {\bibfield
  {journal} {\bibinfo  {journal} {Physical Review Letters}\ }\textbf {\bibinfo
  {volume} {115}},\ \bibinfo {pages} {196403} (\bibinfo {year}
  {2015})}\BibitemShut {NoStop}%
\bibitem [{\citenamefont {Giannozzi}\ \emph {et~al.}(2009)\citenamefont
  {Giannozzi}, \citenamefont {Baroni}, \citenamefont {Bonini}, \citenamefont
  {Calandra}, \citenamefont {Car}, \citenamefont {Cavazzoni}, \citenamefont
  {Ceresoli}, \citenamefont {Chiarotti}, \citenamefont {Cococcioni},
  \citenamefont {Dabo}, \citenamefont {Corso}, \citenamefont {de~Gironcoli},
  \citenamefont {Fabris}, \citenamefont {Fratesi}, \citenamefont {Gebauer},
  \citenamefont {Gerstmann}, \citenamefont {Gougoussis}, \citenamefont
  {Kokalj}, \citenamefont {Lazzeri}, \citenamefont {Martin-Samos},
  \citenamefont {Marzari}, \citenamefont {Mauri}, \citenamefont {Mazzarello},
  \citenamefont {Paolini}, \citenamefont {Pasquarello}, \citenamefont
  {Paulatto}, \citenamefont {Sbraccia}, \citenamefont {Scandalo}, \citenamefont
  {Sclauzero}, \citenamefont {Seitsonen}, \citenamefont {Smogunov},
  \citenamefont {Umari},\ and\ \citenamefont
  {Wentzcovitch}}]{Giannozzi_et_al:2009}%
  \BibitemOpen
  \bibfield  {author} {\bibinfo {author} {\bibfnamefont {P.}~\bibnamefont
  {Giannozzi}}, \bibinfo {author} {\bibfnamefont {S.}~\bibnamefont {Baroni}},
  \bibinfo {author} {\bibfnamefont {N.}~\bibnamefont {Bonini}}, \bibinfo
  {author} {\bibfnamefont {M.}~\bibnamefont {Calandra}}, \bibinfo {author}
  {\bibfnamefont {R.}~\bibnamefont {Car}}, \bibinfo {author} {\bibfnamefont
  {C.}~\bibnamefont {Cavazzoni}}, \bibinfo {author} {\bibfnamefont
  {D.}~\bibnamefont {Ceresoli}}, \bibinfo {author} {\bibfnamefont {G.~L.}\
  \bibnamefont {Chiarotti}}, \bibinfo {author} {\bibfnamefont {M.}~\bibnamefont
  {Cococcioni}}, \bibinfo {author} {\bibfnamefont {I.}~\bibnamefont {Dabo}},
  \bibinfo {author} {\bibfnamefont {A.~D.}\ \bibnamefont {Corso}}, \bibinfo
  {author} {\bibfnamefont {S.}~\bibnamefont {de~Gironcoli}}, \bibinfo {author}
  {\bibfnamefont {S.}~\bibnamefont {Fabris}}, \bibinfo {author} {\bibfnamefont
  {G.}~\bibnamefont {Fratesi}}, \bibinfo {author} {\bibfnamefont
  {R.}~\bibnamefont {Gebauer}}, \bibinfo {author} {\bibfnamefont
  {U.}~\bibnamefont {Gerstmann}}, \bibinfo {author} {\bibfnamefont
  {C.}~\bibnamefont {Gougoussis}}, \bibinfo {author} {\bibfnamefont
  {A.}~\bibnamefont {Kokalj}}, \bibinfo {author} {\bibfnamefont
  {M.}~\bibnamefont {Lazzeri}}, \bibinfo {author} {\bibfnamefont
  {L.}~\bibnamefont {Martin-Samos}}, \bibinfo {author} {\bibfnamefont
  {N.}~\bibnamefont {Marzari}}, \bibinfo {author} {\bibfnamefont
  {F.}~\bibnamefont {Mauri}}, \bibinfo {author} {\bibfnamefont
  {R.}~\bibnamefont {Mazzarello}}, \bibinfo {author} {\bibfnamefont
  {S.}~\bibnamefont {Paolini}}, \bibinfo {author} {\bibfnamefont
  {A.}~\bibnamefont {Pasquarello}}, \bibinfo {author} {\bibfnamefont
  {L.}~\bibnamefont {Paulatto}}, \bibinfo {author} {\bibfnamefont
  {C.}~\bibnamefont {Sbraccia}}, \bibinfo {author} {\bibfnamefont
  {S.}~\bibnamefont {Scandalo}}, \bibinfo {author} {\bibfnamefont
  {G.}~\bibnamefont {Sclauzero}}, \bibinfo {author} {\bibfnamefont {A.~P.}\
  \bibnamefont {Seitsonen}}, \bibinfo {author} {\bibfnamefont {A.}~\bibnamefont
  {Smogunov}}, \bibinfo {author} {\bibfnamefont {P.}~\bibnamefont {Umari}}, \
  and\ \bibinfo {author} {\bibfnamefont {R.}~\bibnamefont {Wentzcovitch}},\
  }\href {\doibase 10.1088/0953-8984/21/39/395502} {\bibfield  {journal}
  {\bibinfo  {journal} {Journal of Physics: Condensed Matter}\ }\textbf
  {\bibinfo {volume} {21}},\ \bibinfo {pages} {395502} (\bibinfo {year}
  {2009})}\BibitemShut {NoStop}%
\bibitem [{\citenamefont {Perdew}\ \emph {et~al.}(1996)\citenamefont {Perdew},
  \citenamefont {Burke},\ and\ \citenamefont {Ernzerhof}}]{Perdew:1996iq}%
  \BibitemOpen
  \bibfield  {author} {\bibinfo {author} {\bibfnamefont {J.~P.}\ \bibnamefont
  {Perdew}}, \bibinfo {author} {\bibfnamefont {K.}~\bibnamefont {Burke}}, \
  and\ \bibinfo {author} {\bibfnamefont {M.}~\bibnamefont {Ernzerhof}},\ }\href
  {\doibase 10.1103/PhysRevLett.77.3865} {\bibfield  {journal} {\bibinfo
  {journal} {Physical Review Letters}\ }\textbf {\bibinfo {volume} {77}},\
  \bibinfo {pages} {3865} (\bibinfo {year} {1996})}\BibitemShut {NoStop}%
\bibitem [{\citenamefont {Bl{\"o}chl}(1994)}]{Blochl:1994dx}%
  \BibitemOpen
  \bibfield  {author} {\bibinfo {author} {\bibfnamefont {P.~E.}\ \bibnamefont
  {Bl{\"o}chl}},\ }\href@noop {} {\bibfield  {journal} {\bibinfo  {journal}
  {Physical Review B}\ }\textbf {\bibinfo {volume} {50}},\ \bibinfo {pages}
  {17953} (\bibinfo {year} {1994})}\BibitemShut {NoStop}%
\bibitem [{\citenamefont {Kresse}\ and\ \citenamefont
  {Hafner}(1993)}]{Kresse:1993bz}%
  \BibitemOpen
  \bibfield  {author} {\bibinfo {author} {\bibfnamefont {G.}~\bibnamefont
  {Kresse}}\ and\ \bibinfo {author} {\bibfnamefont {J.}~\bibnamefont
  {Hafner}},\ }\href {\doibase 10.1103/PhysRevB.47.558} {\bibfield  {journal}
  {\bibinfo  {journal} {Physical Review B}\ }\textbf {\bibinfo {volume} {47}},\
  \bibinfo {pages} {558} (\bibinfo {year} {1993})}\BibitemShut {NoStop}%
\bibitem [{\citenamefont {Kresse}\ and\ \citenamefont
  {Furthm\"uller}(1996)}]{Kresse:1996kl}%
  \BibitemOpen
  \bibfield  {author} {\bibinfo {author} {\bibfnamefont {G.}~\bibnamefont
  {Kresse}}\ and\ \bibinfo {author} {\bibfnamefont {J.}~\bibnamefont
  {Furthm\"uller}},\ }\href {\doibase 10.1103/PhysRevB.54.11169} {\bibfield
  {journal} {\bibinfo  {journal} {Physical Review B}\ }\textbf {\bibinfo
  {volume} {54}},\ \bibinfo {pages} {11169} (\bibinfo {year}
  {1996})}\BibitemShut {NoStop}%
\bibitem [{\citenamefont {Kresse}\ and\ \citenamefont
  {Joubert}(1999)}]{Kresse:1999dk}%
  \BibitemOpen
  \bibfield  {author} {\bibinfo {author} {\bibfnamefont {G.}~\bibnamefont
  {Kresse}}\ and\ \bibinfo {author} {\bibfnamefont {D.}~\bibnamefont
  {Joubert}},\ }\href {\doibase 10.1103/PhysRevB.59.1758} {\bibfield  {journal}
  {\bibinfo  {journal} {Physical Review B}\ }\textbf {\bibinfo {volume} {59}},\
  \bibinfo {pages} {1758} (\bibinfo {year} {1999})}\BibitemShut {NoStop}%
\bibitem [{\citenamefont {Georges}\ \emph {et~al.}(1996)\citenamefont
  {Georges}, \citenamefont {Kotliar}, \citenamefont {Krauth},\ and\
  \citenamefont {Rozenberg}}]{Georges:1996}%
  \BibitemOpen
  \bibfield  {author} {\bibinfo {author} {\bibfnamefont {A.}~\bibnamefont
  {Georges}}, \bibinfo {author} {\bibfnamefont {G.}~\bibnamefont {Kotliar}},
  \bibinfo {author} {\bibfnamefont {W.}~\bibnamefont {Krauth}}, \ and\ \bibinfo
  {author} {\bibfnamefont {M.~J.}\ \bibnamefont {Rozenberg}},\ }\href {\doibase
  10.1103/RevModPhys.68.13} {\bibfield  {journal} {\bibinfo  {journal} {Rev.
  Mod. Phys.}\ }\textbf {\bibinfo {volume} {68}},\ \bibinfo {pages} {13}
  (\bibinfo {year} {1996})}\BibitemShut {NoStop}%
\bibitem [{\citenamefont {Amadon}\ \emph {et~al.}(2008)\citenamefont {Amadon},
  \citenamefont {Lechermann}, \citenamefont {Georges}, \citenamefont {Jollet},
  \citenamefont {Wehling},\ and\ \citenamefont {Lichtenstein}}]{Amadon:2008}%
  \BibitemOpen
  \bibfield  {author} {\bibinfo {author} {\bibfnamefont {B.}~\bibnamefont
  {Amadon}}, \bibinfo {author} {\bibfnamefont {F.}~\bibnamefont {Lechermann}},
  \bibinfo {author} {\bibfnamefont {A.}~\bibnamefont {Georges}}, \bibinfo
  {author} {\bibfnamefont {F.}~\bibnamefont {Jollet}}, \bibinfo {author}
  {\bibfnamefont {T.~O.}\ \bibnamefont {Wehling}}, \ and\ \bibinfo {author}
  {\bibfnamefont {A.~I.}\ \bibnamefont {Lichtenstein}},\ }\href {\doibase
  10.1103/PhysRevB.77.205112} {\bibfield  {journal} {\bibinfo  {journal}
  {Physical Review B}\ }\textbf {\bibinfo {volume} {77}},\ \bibinfo {pages}
  {205112} (\bibinfo {year} {2008})}\BibitemShut {NoStop}%
\bibitem [{\citenamefont {Aichhorn}\ \emph {et~al.}(2016)\citenamefont
  {Aichhorn}, \citenamefont {Pourovskii}, \citenamefont {Seth}, \citenamefont
  {Vildosola}, \citenamefont {Zingl}, \citenamefont {Peil}, \citenamefont
  {Deng}, \citenamefont {Mravlje}, \citenamefont {Kraberger}, \citenamefont
  {Martins}, \citenamefont {Ferrero},\ and\ \citenamefont
  {Parcollet}}]{aichhorn_dfttools_2016}%
  \BibitemOpen
  \bibfield  {author} {\bibinfo {author} {\bibfnamefont {M.}~\bibnamefont
  {Aichhorn}}, \bibinfo {author} {\bibfnamefont {L.}~\bibnamefont
  {Pourovskii}}, \bibinfo {author} {\bibfnamefont {P.}~\bibnamefont {Seth}},
  \bibinfo {author} {\bibfnamefont {V.}~\bibnamefont {Vildosola}}, \bibinfo
  {author} {\bibfnamefont {M.}~\bibnamefont {Zingl}}, \bibinfo {author}
  {\bibfnamefont {O.}~\bibnamefont {Peil}}, \bibinfo {author} {\bibfnamefont
  {X.}~\bibnamefont {Deng}}, \bibinfo {author} {\bibfnamefont {J.}~\bibnamefont
  {Mravlje}}, \bibinfo {author} {\bibfnamefont {G.}~\bibnamefont {Kraberger}},
  \bibinfo {author} {\bibfnamefont {C.}~\bibnamefont {Martins}}, \bibinfo
  {author} {\bibfnamefont {M.}~\bibnamefont {Ferrero}}, \ and\ \bibinfo
  {author} {\bibfnamefont {O.}~\bibnamefont {Parcollet}},\ }\href {\doibase
  doi:10.1016/j.cpc.2016.03.014} {\bibfield  {journal} {\bibinfo  {journal}
  {Computer Physics Communications}\ }\textbf {\bibinfo {volume} {204}},\
  \bibinfo {pages} {200} (\bibinfo {year} {2016})}\BibitemShut {NoStop}%
\bibitem [{\citenamefont {Parcollet}\ \emph {et~al.}(2015)\citenamefont
  {Parcollet}, \citenamefont {Ferrero}, \citenamefont {Ayral}, \citenamefont
  {Hafermann}, \citenamefont {Krivenko}, \citenamefont {Messio},\ and\
  \citenamefont {Seth}}]{parcollet_triqs_2015}%
  \BibitemOpen
  \bibfield  {author} {\bibinfo {author} {\bibfnamefont {O.}~\bibnamefont
  {Parcollet}}, \bibinfo {author} {\bibfnamefont {M.}~\bibnamefont {Ferrero}},
  \bibinfo {author} {\bibfnamefont {T.}~\bibnamefont {Ayral}}, \bibinfo
  {author} {\bibfnamefont {H.}~\bibnamefont {Hafermann}}, \bibinfo {author}
  {\bibfnamefont {I.}~\bibnamefont {Krivenko}}, \bibinfo {author}
  {\bibfnamefont {L.}~\bibnamefont {Messio}}, \ and\ \bibinfo {author}
  {\bibfnamefont {P.}~\bibnamefont {Seth}},\ }\href {\doibase
  https://doi.org/10.1016/j.cpc.2015.04.023} {\bibfield  {journal} {\bibinfo
  {journal} {Computer Physics Communications}\ }\textbf {\bibinfo {volume}
  {196}},\ \bibinfo {pages} {398 } (\bibinfo {year} {2015})}\BibitemShut
  {NoStop}%
\bibitem [{\citenamefont {Gull}\ \emph {et~al.}(2011)\citenamefont {Gull},
  \citenamefont {Millis}, \citenamefont {Lichtenstein}, \citenamefont
  {Rubtsov}, \citenamefont {Troyer},\ and\ \citenamefont {Werner}}]{Gull:2011}%
  \BibitemOpen
  \bibfield  {author} {\bibinfo {author} {\bibfnamefont {E.}~\bibnamefont
  {Gull}}, \bibinfo {author} {\bibfnamefont {A.~J.}\ \bibnamefont {Millis}},
  \bibinfo {author} {\bibfnamefont {A.~I.}\ \bibnamefont {Lichtenstein}},
  \bibinfo {author} {\bibfnamefont {A.~N.}\ \bibnamefont {Rubtsov}}, \bibinfo
  {author} {\bibfnamefont {M.}~\bibnamefont {Troyer}}, \ and\ \bibinfo {author}
  {\bibfnamefont {P.}~\bibnamefont {Werner}},\ }\href {\doibase
  10.1103/RevModPhys.83.349} {\bibfield  {journal} {\bibinfo  {journal} {Rev.
  Mod. Phys.}\ }\textbf {\bibinfo {volume} {83}},\ \bibinfo {pages} {349}
  (\bibinfo {year} {2011})}\BibitemShut {NoStop}%
\bibitem [{\citenamefont {Marzari}\ \emph {et~al.}(2012)\citenamefont
  {Marzari}, \citenamefont {Mostofi}, \citenamefont {Yates}, \citenamefont
  {Souza},\ and\ \citenamefont {Vanderbilt}}]{Marzari_et_al:2012}%
  \BibitemOpen
  \bibfield  {author} {\bibinfo {author} {\bibfnamefont {N.}~\bibnamefont
  {Marzari}}, \bibinfo {author} {\bibfnamefont {A.~A.}\ \bibnamefont
  {Mostofi}}, \bibinfo {author} {\bibfnamefont {J.~R.}\ \bibnamefont {Yates}},
  \bibinfo {author} {\bibfnamefont {I.}~\bibnamefont {Souza}}, \ and\ \bibinfo
  {author} {\bibfnamefont {D.}~\bibnamefont {Vanderbilt}},\ }\href {\doibase
  10.1103/RevModPhys.84.1419} {\bibfield  {journal} {\bibinfo  {journal}
  {Reviews of Modern Physics}\ }\textbf {\bibinfo {volume} {84}},\ \bibinfo
  {pages} {1419} (\bibinfo {year} {2012})}\BibitemShut {NoStop}%
\bibitem [{\citenamefont {Mostofi}\ \emph {et~al.}(2014)\citenamefont
  {Mostofi}, \citenamefont {Yates}, \citenamefont {Pizzi}, \citenamefont {Lee},
  \citenamefont {Souza}, \citenamefont {Vanderbilt},\ and\ \citenamefont
  {Marzari}}]{Mostofi_et_al:2014}%
  \BibitemOpen
  \bibfield  {author} {\bibinfo {author} {\bibfnamefont {A.~A.}\ \bibnamefont
  {Mostofi}}, \bibinfo {author} {\bibfnamefont {J.~R.}\ \bibnamefont {Yates}},
  \bibinfo {author} {\bibfnamefont {G.}~\bibnamefont {Pizzi}}, \bibinfo
  {author} {\bibfnamefont {Y.-S.}\ \bibnamefont {Lee}}, \bibinfo {author}
  {\bibfnamefont {I.}~\bibnamefont {Souza}}, \bibinfo {author} {\bibfnamefont
  {D.}~\bibnamefont {Vanderbilt}}, \ and\ \bibinfo {author} {\bibfnamefont
  {N.}~\bibnamefont {Marzari}},\ }\href {\doibase
  https://doi.org/10.1016/j.cpc.2014.05.003} {\bibfield  {journal} {\bibinfo
  {journal} {Computer Physics Communications}\ }\textbf {\bibinfo {volume}
  {185}},\ \bibinfo {pages} {2309 } (\bibinfo {year} {2014})}\BibitemShut
  {NoStop}%
\bibitem [{\citenamefont {Hampel}\ \emph
  {et~al.}(2019{\natexlab{a}})\citenamefont {Hampel}, \citenamefont {Beck},\
  and\ \citenamefont {Ederer}}]{soliDMFT}%
  \BibitemOpen
  \bibfield  {author} {\bibinfo {author} {\bibfnamefont {A.}~\bibnamefont
  {Hampel}}, \bibinfo {author} {\bibfnamefont {S.}~\bibnamefont {Beck}}, \ and\
  \bibinfo {author} {\bibfnamefont {C.}~\bibnamefont {Ederer}},\ }\href@noop {}
  {\enquote {\bibinfo {title} {{soliDMFT}},}\ }\bibinfo {howpublished}
  {\url{https://github.com/materialstheory/soliDMFT}} (\bibinfo {year}
  {2019}{\natexlab{a}})\BibitemShut {NoStop}%
\bibitem [{\citenamefont {Seth}\ \emph {et~al.}(2016)\citenamefont {Seth},
  \citenamefont {Krivenko}, \citenamefont {Ferrero},\ and\ \citenamefont
  {Parcollet}}]{Seth2016274}%
  \BibitemOpen
  \bibfield  {author} {\bibinfo {author} {\bibfnamefont {P.}~\bibnamefont
  {Seth}}, \bibinfo {author} {\bibfnamefont {I.}~\bibnamefont {Krivenko}},
  \bibinfo {author} {\bibfnamefont {M.}~\bibnamefont {Ferrero}}, \ and\
  \bibinfo {author} {\bibfnamefont {O.}~\bibnamefont {Parcollet}},\ }\href
  {\doibase http://dx.doi.org/10.1016/j.cpc.2015.10.023} {\bibfield  {journal}
  {\bibinfo  {journal} {Computer Physics Communications}\ }\textbf {\bibinfo
  {volume} {200}},\ \bibinfo {pages} {274 } (\bibinfo {year}
  {2016})}\BibitemShut {NoStop}%
\bibitem [{\citenamefont {Vaugier}\ \emph {et~al.}(2012)\citenamefont
  {Vaugier}, \citenamefont {Jiang},\ and\ \citenamefont
  {Biermann}}]{vaugier2012}%
  \BibitemOpen
  \bibfield  {author} {\bibinfo {author} {\bibfnamefont {L.}~\bibnamefont
  {Vaugier}}, \bibinfo {author} {\bibfnamefont {H.}~\bibnamefont {Jiang}}, \
  and\ \bibinfo {author} {\bibfnamefont {S.}~\bibnamefont {Biermann}},\ }\href
  {\doibase 10.1103/PhysRevB.86.165105} {\bibfield  {journal} {\bibinfo
  {journal} {Physical Review B}\ }\textbf {\bibinfo {volume} {86}},\ \bibinfo
  {pages} {165105} (\bibinfo {year} {2012})}\BibitemShut {NoStop}%
\bibitem [{\citenamefont {Boehnke}\ \emph {et~al.}(2011)\citenamefont
  {Boehnke}, \citenamefont {Hafermann}, \citenamefont {Ferrero}, \citenamefont
  {Lechermann},\ and\ \citenamefont {Parcollet}}]{boehnke:2011}%
  \BibitemOpen
  \bibfield  {author} {\bibinfo {author} {\bibfnamefont {L.}~\bibnamefont
  {Boehnke}}, \bibinfo {author} {\bibfnamefont {H.}~\bibnamefont {Hafermann}},
  \bibinfo {author} {\bibfnamefont {M.}~\bibnamefont {Ferrero}}, \bibinfo
  {author} {\bibfnamefont {F.}~\bibnamefont {Lechermann}}, \ and\ \bibinfo
  {author} {\bibfnamefont {O.}~\bibnamefont {Parcollet}},\ }\href {\doibase
  10.1103/PhysRevB.84.075145} {\bibfield  {journal} {\bibinfo  {journal}
  {Physical Review B}\ }\textbf {\bibinfo {volume} {84}},\ \bibinfo {pages}
  {075145} (\bibinfo {year} {2011})}\BibitemShut {NoStop}%
\bibitem [{\citenamefont {Solovyev}\ \emph {et~al.}(1994)\citenamefont
  {Solovyev}, \citenamefont {Dederichs},\ and\ \citenamefont
  {Anisimov}}]{Solovyev:1994}%
  \BibitemOpen
  \bibfield  {author} {\bibinfo {author} {\bibfnamefont {I.~V.}\ \bibnamefont
  {Solovyev}}, \bibinfo {author} {\bibfnamefont {P.~H.}\ \bibnamefont
  {Dederichs}}, \ and\ \bibinfo {author} {\bibfnamefont {V.~I.}\ \bibnamefont
  {Anisimov}},\ }\href {\doibase 10.1103/PhysRevB.50.16861} {\bibfield
  {journal} {\bibinfo  {journal} {Phys. Rev. B}\ }\textbf {\bibinfo {volume}
  {50}},\ \bibinfo {pages} {16861} (\bibinfo {year} {1994})}\BibitemShut
  {NoStop}%
\bibitem [{\citenamefont {Anisimov}\ \emph {et~al.}(1997)\citenamefont
  {Anisimov}, \citenamefont {Aryasetiawan},\ and\ \citenamefont
  {Lichtenstein}}]{anisimov1997}%
  \BibitemOpen
  \bibfield  {author} {\bibinfo {author} {\bibfnamefont {V.~I.}\ \bibnamefont
  {Anisimov}}, \bibinfo {author} {\bibfnamefont {F.}~\bibnamefont
  {Aryasetiawan}}, \ and\ \bibinfo {author} {\bibfnamefont {A.~I.}\
  \bibnamefont {Lichtenstein}},\ }\href
  {http://stacks.iop.org/0953-8984/9/i=4/a=002} {\bibfield  {journal} {\bibinfo
   {journal} {Journal of Physics: Condensed Matter}\ }\textbf {\bibinfo
  {volume} {9}},\ \bibinfo {pages} {767} (\bibinfo {year} {1997})}\BibitemShut
  {NoStop}%
\bibitem [{\citenamefont {Fuchs}\ \emph {et~al.}(2011)\citenamefont {Fuchs},
  \citenamefont {Gull}, \citenamefont {Troyer}, \citenamefont {Jarrell},\ and\
  \citenamefont {Pruschke}}]{Fuchs:2011}%
  \BibitemOpen
  \bibfield  {author} {\bibinfo {author} {\bibfnamefont {S.}~\bibnamefont
  {Fuchs}}, \bibinfo {author} {\bibfnamefont {E.}~\bibnamefont {Gull}},
  \bibinfo {author} {\bibfnamefont {M.}~\bibnamefont {Troyer}}, \bibinfo
  {author} {\bibfnamefont {M.}~\bibnamefont {Jarrell}}, \ and\ \bibinfo
  {author} {\bibfnamefont {T.}~\bibnamefont {Pruschke}},\ }\href {\doibase
  10.1103/PhysRevB.83.235113} {\bibfield  {journal} {\bibinfo  {journal}
  {Physical Review B}\ }\textbf {\bibinfo {volume} {83}},\ \bibinfo {pages}
  {235113} (\bibinfo {year} {2011})}\BibitemShut {NoStop}%
\bibitem [{\citenamefont {Jarrell}\ and\ \citenamefont
  {Gubernatis}(1996)}]{Jarrel:2010}%
  \BibitemOpen
  \bibfield  {author} {\bibinfo {author} {\bibfnamefont {M.}~\bibnamefont
  {Jarrell}}\ and\ \bibinfo {author} {\bibfnamefont {J.}~\bibnamefont
  {Gubernatis}},\ }\href {\doibase
  https://doi.org/10.1016/0370-1573(95)00074-7} {\bibfield  {journal} {\bibinfo
   {journal} {Physics Reports}\ }\textbf {\bibinfo {volume} {269}},\ \bibinfo
  {pages} {133 } (\bibinfo {year} {1996})}\BibitemShut {NoStop}%
\bibitem [{\citenamefont {Abrikosov}\ \emph {et~al.}(2012)\citenamefont
  {Abrikosov}, \citenamefont {{Gorkov, L.P.}}, \citenamefont {Dzyaloshinski},\
  and\ \citenamefont {Silverman}}]{abrikosov2012methods}%
  \BibitemOpen
  \bibfield  {author} {\bibinfo {author} {\bibfnamefont {A.~A.}\ \bibnamefont
  {Abrikosov}}, \bibinfo {author} {\bibnamefont {{Gorkov, L.P.}}}, \bibinfo
  {author} {\bibfnamefont {I.~E.}\ \bibnamefont {Dzyaloshinski}}, \ and\
  \bibinfo {author} {\bibfnamefont {R.~A.}\ \bibnamefont {Silverman}},\
  }\href@noop {} {\emph {\bibinfo {title} {{Methods of Quantum Field Theory in
  Statistical Physics}}}},\ Dover Books on Physics\ (\bibinfo  {publisher}
  {Dover Publications},\ \bibinfo {year} {2012})\BibitemShut {NoStop}%
\bibitem [{\citenamefont {Galitskii}\ and\ \citenamefont
  {Migdal}(1958)}]{galitskii1958}%
  \BibitemOpen
  \bibfield  {author} {\bibinfo {author} {\bibfnamefont {V.}~\bibnamefont
  {Galitskii}}\ and\ \bibinfo {author} {\bibfnamefont {A.}~\bibnamefont
  {Migdal}},\ }\href@noop {} {\bibfield  {journal} {\bibinfo  {journal} {Soviet
  Physics Journal of Experimental and Theoretical Physics}\ }\textbf {\bibinfo
  {volume} {7}},\ \bibinfo {pages} {96} (\bibinfo {year} {1958})}\BibitemShut
  {NoStop}%
\bibitem [{\citenamefont {Nekrasov}\ \emph {et~al.}(2005)\citenamefont
  {Nekrasov}, \citenamefont {Keller}, \citenamefont {Kondakov}, \citenamefont
  {Kozhevnikov}, \citenamefont {Pruschke}, \citenamefont {Held}, \citenamefont
  {Vollhardt},\ and\ \citenamefont {Anisimov}}]{Nekrasov:2005}%
  \BibitemOpen
  \bibfield  {author} {\bibinfo {author} {\bibfnamefont {I.~A.}\ \bibnamefont
  {Nekrasov}}, \bibinfo {author} {\bibfnamefont {G.}~\bibnamefont {Keller}},
  \bibinfo {author} {\bibfnamefont {D.~E.}\ \bibnamefont {Kondakov}}, \bibinfo
  {author} {\bibfnamefont {A.~V.}\ \bibnamefont {Kozhevnikov}}, \bibinfo
  {author} {\bibfnamefont {T.}~\bibnamefont {Pruschke}}, \bibinfo {author}
  {\bibfnamefont {K.}~\bibnamefont {Held}}, \bibinfo {author} {\bibfnamefont
  {D.}~\bibnamefont {Vollhardt}}, \ and\ \bibinfo {author} {\bibfnamefont
  {V.~I.}\ \bibnamefont {Anisimov}},\ }\href {\doibase
  10.1103/PhysRevB.72.155106} {\bibfield  {journal} {\bibinfo  {journal}
  {Physical Review B}\ }\textbf {\bibinfo {volume} {72}},\ \bibinfo {pages}
  {155106} (\bibinfo {year} {2005})}\BibitemShut {NoStop}%
\bibitem [{\citenamefont {Pavarini}\ \emph {et~al.}(2004)\citenamefont
  {Pavarini}, \citenamefont {Biermann}, \citenamefont {Poteryaev},
  \citenamefont {Lichtenstein}, \citenamefont {Georges},\ and\ \citenamefont
  {Andersen}}]{Pavarini_et_al:2004}%
  \BibitemOpen
  \bibfield  {author} {\bibinfo {author} {\bibfnamefont {E.}~\bibnamefont
  {Pavarini}}, \bibinfo {author} {\bibfnamefont {S.}~\bibnamefont {Biermann}},
  \bibinfo {author} {\bibfnamefont {A.}~\bibnamefont {Poteryaev}}, \bibinfo
  {author} {\bibfnamefont {A.~I.}\ \bibnamefont {Lichtenstein}}, \bibinfo
  {author} {\bibfnamefont {A.}~\bibnamefont {Georges}}, \ and\ \bibinfo
  {author} {\bibfnamefont {O.~K.}\ \bibnamefont {Andersen}},\ }\href {\doibase
  10.1103/PhysRevLett.92.176403} {\bibfield  {journal} {\bibinfo  {journal}
  {Physical Review Letters}\ }\textbf {\bibinfo {volume} {92}},\ \bibinfo
  {pages} {176403} (\bibinfo {year} {2004})}\BibitemShut {NoStop}%
\bibitem [{\citenamefont {Park}\ \emph {et~al.}(2012)\citenamefont {Park},
  \citenamefont {Millis},\ and\ \citenamefont {Marianetti}}]{Park:2012hg}%
  \BibitemOpen
  \bibfield  {author} {\bibinfo {author} {\bibfnamefont {H.}~\bibnamefont
  {Park}}, \bibinfo {author} {\bibfnamefont {A.~J.}\ \bibnamefont {Millis}}, \
  and\ \bibinfo {author} {\bibfnamefont {C.~A.}\ \bibnamefont {Marianetti}},\
  }\href {\doibase 10.1103/PhysRevLett.109.156402} {\bibfield  {journal}
  {\bibinfo  {journal} {Physical Review Letters}\ }\textbf {\bibinfo {volume}
  {109}},\ \bibinfo {pages} {156402} (\bibinfo {year} {2012})}\BibitemShut
  {NoStop}%
\bibitem [{\citenamefont {Subedi}\ \emph {et~al.}(2015)\citenamefont {Subedi},
  \citenamefont {Peil},\ and\ \citenamefont {Georges}}]{Subedi:2015en}%
  \BibitemOpen
  \bibfield  {author} {\bibinfo {author} {\bibfnamefont {A.}~\bibnamefont
  {Subedi}}, \bibinfo {author} {\bibfnamefont {O.~E.}\ \bibnamefont {Peil}}, \
  and\ \bibinfo {author} {\bibfnamefont {A.}~\bibnamefont {Georges}},\
  }\href@noop {} {\bibfield  {journal} {\bibinfo  {journal} {Physical Review
  B}\ }\textbf {\bibinfo {volume} {91}},\ \bibinfo {pages} {075128} (\bibinfo
  {year} {2015})}\BibitemShut {NoStop}%
\bibitem [{\citenamefont {McNally}\ \emph {et~al.}(2019)\citenamefont
  {McNally}, \citenamefont {Lu}, \citenamefont {Pelliciari}, \citenamefont
  {Beck}, \citenamefont {Dantz}, \citenamefont {Naamneh}, \citenamefont
  {Shang}, \citenamefont {Medarde}, \citenamefont {Schneider}, \citenamefont
  {Strocov} \emph {et~al.}}]{Mcnally:2019}%
  \BibitemOpen
  \bibfield  {author} {\bibinfo {author} {\bibfnamefont {D.~E.}\ \bibnamefont
  {McNally}}, \bibinfo {author} {\bibfnamefont {X.}~\bibnamefont {Lu}},
  \bibinfo {author} {\bibfnamefont {J.}~\bibnamefont {Pelliciari}}, \bibinfo
  {author} {\bibfnamefont {S.}~\bibnamefont {Beck}}, \bibinfo {author}
  {\bibfnamefont {M.}~\bibnamefont {Dantz}}, \bibinfo {author} {\bibfnamefont
  {M.}~\bibnamefont {Naamneh}}, \bibinfo {author} {\bibfnamefont
  {T.}~\bibnamefont {Shang}}, \bibinfo {author} {\bibfnamefont
  {M.}~\bibnamefont {Medarde}}, \bibinfo {author} {\bibfnamefont {C.~W.}\
  \bibnamefont {Schneider}}, \bibinfo {author} {\bibfnamefont {V.~N.}\
  \bibnamefont {Strocov}},  \emph {et~al.},\ }\href@noop {} {\bibfield
  {journal} {\bibinfo  {journal} {npj Quantum Materials}\ }\textbf {\bibinfo
  {volume} {4}},\ \bibinfo {pages} {6} (\bibinfo {year} {2019})}\BibitemShut
  {NoStop}%
\bibitem [{\citenamefont {Pavarini}\ \emph {et~al.}(2005)\citenamefont
  {Pavarini}, \citenamefont {Yamasaki}, \citenamefont {Nuss},\ and\
  \citenamefont {Andersen}}]{Pavarini:2005}%
  \BibitemOpen
  \bibfield  {author} {\bibinfo {author} {\bibfnamefont {E.}~\bibnamefont
  {Pavarini}}, \bibinfo {author} {\bibfnamefont {A.}~\bibnamefont {Yamasaki}},
  \bibinfo {author} {\bibfnamefont {J.}~\bibnamefont {Nuss}}, \ and\ \bibinfo
  {author} {\bibfnamefont {O.~K.}\ \bibnamefont {Andersen}},\ }\href {\doibase
  10.1088/1367-2630/7/1/188} {\bibfield  {journal} {\bibinfo  {journal} {New
  Journal of Physics}\ }\textbf {\bibinfo {volume} {7}},\ \bibinfo {pages}
  {188} (\bibinfo {year} {2005})}\BibitemShut {NoStop}%
\bibitem [{\citenamefont {Sclauzero}\ \emph {et~al.}(2016)\citenamefont
  {Sclauzero}, \citenamefont {Dymkowski},\ and\ \citenamefont
  {Ederer}}]{Sclauzero/Dymkowski/Ederer:2016}%
  \BibitemOpen
  \bibfield  {author} {\bibinfo {author} {\bibfnamefont {G.}~\bibnamefont
  {Sclauzero}}, \bibinfo {author} {\bibfnamefont {K.}~\bibnamefont
  {Dymkowski}}, \ and\ \bibinfo {author} {\bibfnamefont {C.}~\bibnamefont
  {Ederer}},\ }\href {\doibase 10.1103/PhysRevB.94.245109} {\bibfield
  {journal} {\bibinfo  {journal} {Physical Review B}\ }\textbf {\bibinfo
  {volume} {94}},\ \bibinfo {pages} {245109} (\bibinfo {year}
  {2016})}\BibitemShut {NoStop}%
\bibitem [{\citenamefont {Medarde}\ \emph {et~al.}(2008)\citenamefont
  {Medarde}, \citenamefont {Fern\'andez-D\'{\i}az},\ and\ \citenamefont
  {Lacorre}}]{Medarde2008}%
  \BibitemOpen
  \bibfield  {author} {\bibinfo {author} {\bibfnamefont {M.}~\bibnamefont
  {Medarde}}, \bibinfo {author} {\bibfnamefont {M.~T.}\ \bibnamefont
  {Fern\'andez-D\'{\i}az}}, \ and\ \bibinfo {author} {\bibfnamefont
  {P.}~\bibnamefont {Lacorre}},\ }\href {\doibase 10.1103/PhysRevB.78.212101}
  {\bibfield  {journal} {\bibinfo  {journal} {Physical Review B}\ }\textbf
  {\bibinfo {volume} {78}},\ \bibinfo {pages} {212101} (\bibinfo {year}
  {2008})}\BibitemShut {NoStop}%
\bibitem [{\citenamefont {Peil}\ \emph {et~al.}(2019)\citenamefont {Peil},
  \citenamefont {Hampel}, \citenamefont {Ederer},\ and\ \citenamefont
  {Georges}}]{peil:2019}%
  \BibitemOpen
  \bibfield  {author} {\bibinfo {author} {\bibfnamefont {O.~E.}\ \bibnamefont
  {Peil}}, \bibinfo {author} {\bibfnamefont {A.}~\bibnamefont {Hampel}},
  \bibinfo {author} {\bibfnamefont {C.}~\bibnamefont {Ederer}}, \ and\ \bibinfo
  {author} {\bibfnamefont {A.}~\bibnamefont {Georges}},\ }\href {\doibase
  10.1103/PhysRevB.99.245127} {\bibfield  {journal} {\bibinfo  {journal}
  {Physical Review B}\ }\textbf {\bibinfo {volume} {99}},\ \bibinfo {pages}
  {245127} (\bibinfo {year} {2019})}\BibitemShut {NoStop}%
\bibitem [{\citenamefont {Hampel}\ \emph
  {et~al.}(2019{\natexlab{b}})\citenamefont {Hampel}, \citenamefont {Liu},
  \citenamefont {Franchini},\ and\ \citenamefont {Ederer}}]{hampel:2019}%
  \BibitemOpen
  \bibfield  {author} {\bibinfo {author} {\bibfnamefont {A.}~\bibnamefont
  {Hampel}}, \bibinfo {author} {\bibfnamefont {P.}~\bibnamefont {Liu}},
  \bibinfo {author} {\bibfnamefont {C.}~\bibnamefont {Franchini}}, \ and\
  \bibinfo {author} {\bibfnamefont {C.}~\bibnamefont {Ederer}},\ }\href
  {\doibase 10.1038/s41535-019-0145-4} {\bibfield  {journal} {\bibinfo
  {journal} {npj Quantum Materials}\ }\textbf {\bibinfo {volume} {4}},\
  \bibinfo {pages} {5} (\bibinfo {year} {2019}{\natexlab{b}})}\BibitemShut
  {NoStop}%
\bibitem [{\citenamefont {Mercy}\ \emph {et~al.}(2017)\citenamefont {Mercy},
  \citenamefont {Bieder}, \citenamefont {{\'I}{\~n}iguez},\ and\ \citenamefont
  {Ghosez}}]{Mercy2017}%
  \BibitemOpen
  \bibfield  {author} {\bibinfo {author} {\bibfnamefont {A.}~\bibnamefont
  {Mercy}}, \bibinfo {author} {\bibfnamefont {J.}~\bibnamefont {Bieder}},
  \bibinfo {author} {\bibfnamefont {J.}~\bibnamefont {{\'I}{\~n}iguez}}, \ and\
  \bibinfo {author} {\bibfnamefont {P.}~\bibnamefont {Ghosez}},\ }\href
  {\doibase 10.1038/s41467-017-01811-x} {\bibfield  {journal} {\bibinfo
  {journal} {Nature Communications}\ }\textbf {\bibinfo {volume} {8}},\
  \bibinfo {pages} {1677} (\bibinfo {year} {2017})}\BibitemShut {NoStop}%
\bibitem [{\citenamefont {Varignon}\ \emph {et~al.}(2017)\citenamefont
  {Varignon}, \citenamefont {Grisolia}, \citenamefont {{\'I}{\~n}iguez},
  \citenamefont {Barth{\'e}l{\'e}my},\ and\ \citenamefont
  {Bibes}}]{Varignon:2017is}%
  \BibitemOpen
  \bibfield  {author} {\bibinfo {author} {\bibfnamefont {J.}~\bibnamefont
  {Varignon}}, \bibinfo {author} {\bibfnamefont {M.~N.}\ \bibnamefont
  {Grisolia}}, \bibinfo {author} {\bibfnamefont {J.}~\bibnamefont
  {{\'I}{\~n}iguez}}, \bibinfo {author} {\bibfnamefont {A.}~\bibnamefont
  {Barth{\'e}l{\'e}my}}, \ and\ \bibinfo {author} {\bibfnamefont
  {M.}~\bibnamefont {Bibes}},\ }\href@noop {} {\bibfield  {journal} {\bibinfo
  {journal} {Nature Partner Journals Quantum Materials}\ }\textbf {\bibinfo
  {volume} {2}},\ \bibinfo {pages} {21} (\bibinfo {year} {2017})}\BibitemShut
  {NoStop}%
\bibitem [{\citenamefont {Hampel}\ and\ \citenamefont
  {Ederer}(2017)}]{hampel2017}%
  \BibitemOpen
  \bibfield  {author} {\bibinfo {author} {\bibfnamefont {A.}~\bibnamefont
  {Hampel}}\ and\ \bibinfo {author} {\bibfnamefont {C.}~\bibnamefont
  {Ederer}},\ }\href {\doibase 10.1103/PhysRevB.96.165130} {\bibfield
  {journal} {\bibinfo  {journal} {Physical Review B}\ }\textbf {\bibinfo
  {volume} {96}},\ \bibinfo {pages} {165130} (\bibinfo {year}
  {2017})}\BibitemShut {NoStop}%
\bibitem [{\citenamefont {Park}\ \emph
  {et~al.}(2014{\natexlab{b}})\citenamefont {Park}, \citenamefont {Millis},\
  and\ \citenamefont {Marianetti}}]{Park2014short}%
  \BibitemOpen
  \bibfield  {author} {\bibinfo {author} {\bibfnamefont {H.}~\bibnamefont
  {Park}}, \bibinfo {author} {\bibfnamefont {A.~J.}\ \bibnamefont {Millis}}, \
  and\ \bibinfo {author} {\bibfnamefont {C.~A.}\ \bibnamefont {Marianetti}},\
  }\href {\doibase 10.1103/PhysRevB.89.245133} {\bibfield  {journal} {\bibinfo
  {journal} {Physical Review B}\ }\textbf {\bibinfo {volume} {89}},\ \bibinfo
  {pages} {245133} (\bibinfo {year} {2014}{\natexlab{b}})}\BibitemShut
  {NoStop}%
\bibitem [{\citenamefont {Mazin}\ \emph {et~al.}(2007)\citenamefont {Mazin},
  \citenamefont {Khomskii}, \citenamefont {Lengsdorf}, \citenamefont {Alonso},
  \citenamefont {Marshall}, \citenamefont {Ibberson}, \citenamefont
  {Podlesnyak}, \citenamefont {Mart\'{\i}nez-Lope},\ and\ \citenamefont
  {Abd-Elmeguid}}]{Mazin:2007}%
  \BibitemOpen
  \bibfield  {author} {\bibinfo {author} {\bibfnamefont {I.~I.}\ \bibnamefont
  {Mazin}}, \bibinfo {author} {\bibfnamefont {D.~I.}\ \bibnamefont {Khomskii}},
  \bibinfo {author} {\bibfnamefont {R.}~\bibnamefont {Lengsdorf}}, \bibinfo
  {author} {\bibfnamefont {J.~A.}\ \bibnamefont {Alonso}}, \bibinfo {author}
  {\bibfnamefont {W.~G.}\ \bibnamefont {Marshall}}, \bibinfo {author}
  {\bibfnamefont {R.~M.}\ \bibnamefont {Ibberson}}, \bibinfo {author}
  {\bibfnamefont {A.}~\bibnamefont {Podlesnyak}}, \bibinfo {author}
  {\bibfnamefont {M.~J.}\ \bibnamefont {Mart\'{\i}nez-Lope}}, \ and\ \bibinfo
  {author} {\bibfnamefont {M.~M.}\ \bibnamefont {Abd-Elmeguid}},\ }\href
  {\doibase 10.1103/PhysRevLett.98.176406} {\bibfield  {journal} {\bibinfo
  {journal} {Physical Review Letters}\ }\textbf {\bibinfo {volume} {98}},\
  \bibinfo {pages} {176406} (\bibinfo {year} {2007})}\BibitemShut {NoStop}%
\bibitem [{\citenamefont {Seth}\ \emph
  {et~al.}(2017{\natexlab{a}})\citenamefont {Seth}, \citenamefont {Hansmann},
  \citenamefont {van Roekeghem}, \citenamefont {Vaugier},\ and\ \citenamefont
  {Biermann}}]{Seth:2017}%
  \BibitemOpen
  \bibfield  {author} {\bibinfo {author} {\bibfnamefont {P.}~\bibnamefont
  {Seth}}, \bibinfo {author} {\bibfnamefont {P.}~\bibnamefont {Hansmann}},
  \bibinfo {author} {\bibfnamefont {A.}~\bibnamefont {van Roekeghem}}, \bibinfo
  {author} {\bibfnamefont {L.}~\bibnamefont {Vaugier}}, \ and\ \bibinfo
  {author} {\bibfnamefont {S.}~\bibnamefont {Biermann}},\ }\href {\doibase
  10.1103/PhysRevLett.119.056401} {\bibfield  {journal} {\bibinfo  {journal}
  {Physical Review Letters}\ }\textbf {\bibinfo {volume} {119}},\ \bibinfo
  {pages} {056401} (\bibinfo {year} {2017}{\natexlab{a}})}\BibitemShut
  {NoStop}%
\bibitem [{\citenamefont {Isidori}\ \emph {et~al.}(2019)\citenamefont
  {Isidori}, \citenamefont {Berovi\ifmmode~\acute{c}\else \'{c}\fi{}},
  \citenamefont {Fanfarillo}, \citenamefont {de' Medici}, \citenamefont
  {Fabrizio},\ and\ \citenamefont {Capone}}]{Isidori:2019}%
  \BibitemOpen
  \bibfield  {author} {\bibinfo {author} {\bibfnamefont {A.}~\bibnamefont
  {Isidori}}, \bibinfo {author} {\bibfnamefont {M.}~\bibnamefont
  {Berovi\ifmmode~\acute{c}\else \'{c}\fi{}}}, \bibinfo {author} {\bibfnamefont
  {L.}~\bibnamefont {Fanfarillo}}, \bibinfo {author} {\bibfnamefont
  {L.}~\bibnamefont {de' Medici}}, \bibinfo {author} {\bibfnamefont
  {M.}~\bibnamefont {Fabrizio}}, \ and\ \bibinfo {author} {\bibfnamefont
  {M.}~\bibnamefont {Capone}},\ }\href {\doibase
  10.1103/PhysRevLett.122.186401} {\bibfield  {journal} {\bibinfo  {journal}
  {Physical Review Letters}\ }\textbf {\bibinfo {volume} {122}},\ \bibinfo
  {pages} {186401} (\bibinfo {year} {2019})}\BibitemShut {NoStop}%
\bibitem [{\citenamefont {Perez-Mato}\ \emph {et~al.}(2010)\citenamefont
  {Perez-Mato}, \citenamefont {Orobengoa},\ and\ \citenamefont
  {Aroyo}}]{PerezMato:2010ix}%
  \BibitemOpen
  \bibfield  {author} {\bibinfo {author} {\bibfnamefont {J.~M.}\ \bibnamefont
  {Perez-Mato}}, \bibinfo {author} {\bibfnamefont {D.}~\bibnamefont
  {Orobengoa}}, \ and\ \bibinfo {author} {\bibfnamefont {M.~I.}\ \bibnamefont
  {Aroyo}},\ }\href@noop {} {\bibfield  {journal} {\bibinfo  {journal} {Acta
  Crystallographica A}\ }\textbf {\bibinfo {volume} {66}},\ \bibinfo {pages}
  {558} (\bibinfo {year} {2010})}\BibitemShut {NoStop}%
\bibitem [{\citenamefont {Balachandran}\ and\ \citenamefont
  {Rondinelli}(2013)}]{Balachandran:2013cg}%
  \BibitemOpen
  \bibfield  {author} {\bibinfo {author} {\bibfnamefont {P.~V.}\ \bibnamefont
  {Balachandran}}\ and\ \bibinfo {author} {\bibfnamefont {J.~M.}\ \bibnamefont
  {Rondinelli}},\ }\href {\doibase 10.1103/PhysRevB.88.054101} {\bibfield
  {journal} {\bibinfo  {journal} {Physical Review B}\ }\textbf {\bibinfo
  {volume} {88}},\ \bibinfo {pages} {054101} (\bibinfo {year}
  {2013})}\BibitemShut {NoStop}%
\bibitem [{\citenamefont {Campbell}\ \emph {et~al.}(2006)\citenamefont
  {Campbell}, \citenamefont {Stokes}, \citenamefont {Tanner},\ and\
  \citenamefont {Hatch}}]{Campbell:2006}%
  \BibitemOpen
  \bibfield  {author} {\bibinfo {author} {\bibfnamefont {B.~J.}\ \bibnamefont
  {Campbell}}, \bibinfo {author} {\bibfnamefont {H.~T.}\ \bibnamefont
  {Stokes}}, \bibinfo {author} {\bibfnamefont {D.~E.}\ \bibnamefont {Tanner}},
  \ and\ \bibinfo {author} {\bibfnamefont {D.~M.}\ \bibnamefont {Hatch}},\
  }\href@noop {} {\bibfield  {journal} {\bibinfo  {journal} {Journal of Applied
  Crystallography}\ }\textbf {\bibinfo {volume} {39}},\ \bibinfo {pages} {607}
  (\bibinfo {year} {2006})}\BibitemShut {NoStop}%
\bibitem [{\citenamefont {Alonso}\ \emph {et~al.}(2001)\citenamefont {Alonso},
  \citenamefont {Mart\'{\i}nez-Lope}, \citenamefont {Casais}, \citenamefont
  {Garc\'{\i}a-Mu\~noz}, \citenamefont {Fern\'andez-D\'{\i}az},\ and\
  \citenamefont {Aranda}}]{Alonso:2001bs}%
  \BibitemOpen
  \bibfield  {author} {\bibinfo {author} {\bibfnamefont {J.~A.}\ \bibnamefont
  {Alonso}}, \bibinfo {author} {\bibfnamefont {M.~J.}\ \bibnamefont
  {Mart\'{\i}nez-Lope}}, \bibinfo {author} {\bibfnamefont {M.~T.}\ \bibnamefont
  {Casais}}, \bibinfo {author} {\bibfnamefont {J.~L.}\ \bibnamefont
  {Garc\'{\i}a-Mu\~noz}}, \bibinfo {author} {\bibfnamefont {M.~T.}\
  \bibnamefont {Fern\'andez-D\'{\i}az}}, \ and\ \bibinfo {author}
  {\bibfnamefont {M.~A.~G.}\ \bibnamefont {Aranda}},\ }\href {\doibase
  10.1103/PhysRevB.64.094102} {\bibfield  {journal} {\bibinfo  {journal}
  {Physical Review B}\ }\textbf {\bibinfo {volume} {64}},\ \bibinfo {pages}
  {094102} (\bibinfo {year} {2001})}\BibitemShut {NoStop}%
\bibitem [{\citenamefont {Seth}\ \emph
  {et~al.}(2017{\natexlab{b}})\citenamefont {Seth}, \citenamefont {Peil},
  \citenamefont {Pourovskii}, \citenamefont {Betzinger}, \citenamefont
  {Friedrich}, \citenamefont {Parcollet}, \citenamefont {Biermann},
  \citenamefont {Aryasetiawan},\ and\ \citenamefont
  {Georges}}]{Seth_Georges:2017}%
  \BibitemOpen
  \bibfield  {author} {\bibinfo {author} {\bibfnamefont {P.}~\bibnamefont
  {Seth}}, \bibinfo {author} {\bibfnamefont {O.~E.}\ \bibnamefont {Peil}},
  \bibinfo {author} {\bibfnamefont {L.}~\bibnamefont {Pourovskii}}, \bibinfo
  {author} {\bibfnamefont {M.}~\bibnamefont {Betzinger}}, \bibinfo {author}
  {\bibfnamefont {C.}~\bibnamefont {Friedrich}}, \bibinfo {author}
  {\bibfnamefont {O.}~\bibnamefont {Parcollet}}, \bibinfo {author}
  {\bibfnamefont {S.}~\bibnamefont {Biermann}}, \bibinfo {author}
  {\bibfnamefont {F.}~\bibnamefont {Aryasetiawan}}, \ and\ \bibinfo {author}
  {\bibfnamefont {A.}~\bibnamefont {Georges}},\ }\href {\doibase
  10.1103/PhysRevB.96.205139} {\bibfield  {journal} {\bibinfo  {journal}
  {Physical Review B}\ }\textbf {\bibinfo {volume} {96}},\ \bibinfo {pages}
  {205139} (\bibinfo {year} {2017}{\natexlab{b}})}\BibitemShut {NoStop}%
\bibitem [{\citenamefont {Pourovskii}\ \emph {et~al.}(2007)\citenamefont
  {Pourovskii}, \citenamefont {Amadon}, \citenamefont {Biermann},\ and\
  \citenamefont {Georges}}]{PhysRevB.76.235101}%
  \BibitemOpen
  \bibfield  {author} {\bibinfo {author} {\bibfnamefont {L.~V.}\ \bibnamefont
  {Pourovskii}}, \bibinfo {author} {\bibfnamefont {B.}~\bibnamefont {Amadon}},
  \bibinfo {author} {\bibfnamefont {S.}~\bibnamefont {Biermann}}, \ and\
  \bibinfo {author} {\bibfnamefont {A.}~\bibnamefont {Georges}},\ }\href
  {\doibase 10.1103/PhysRevB.76.235101} {\bibfield  {journal} {\bibinfo
  {journal} {Physical Review B}\ }\textbf {\bibinfo {volume} {76}},\ \bibinfo
  {pages} {235101} (\bibinfo {year} {2007})}\BibitemShut {NoStop}%
\end{thebibliography}%

\end{document}